\documentclass[10pt]{article} 
\usepackage[preprint]{tmlr}

\usepackage{graphicx}%
\usepackage{subcaption}
\usepackage{multirow}%
\usepackage{amsmath,amssymb,amsfonts}%
\usepackage{amsthm}%
\usepackage{mathrsfs}%
\usepackage[title]{appendix}%
\usepackage[table,xcdraw, dvipsnames]{xcolor}
\usepackage{textcomp}%
\usepackage{manyfoot}%
\usepackage{booktabs}%
\usepackage{listings}%
\usepackage{titlesec}
\usepackage{xr}
\usepackage{xr-hyper}
\usepackage{hyperref}
\usepackage{lineno}

\hypersetup{
	hidelinks,
	colorlinks=true,
	linkcolor=Blue,
	filecolor=Blue,
	citecolor=Blue,
	urlcolor=Blue,
	breaklinks=false
}

\makeatletter
\newcommand*{\addFileDependency}[1]{
	\typeout{(#1)}
	\@addtofilelist{#1}
	\IfFileExists{#1}{\typeout{File #1 O.K.}}{\typeout{No file #1.}}
}
\makeatother

\usepackage{hyperref}
\usepackage{url}

\title{Illumination Angular Spectrum Encoding for Controlling the Functionality of Diffractive Networks}

\author{\name Matan Kleiner \email matankleiner@campus.technion.ac.il \\
      \addr Faculty of Electrical and Computer Engineering\\
      Technion, 32000 Haifa, Israel
      \AND
      \name Lior Michaeli \email liormic1@tauex.tau.ac.il \\
      \addr School of Electrical and Computer Engineering, Faculty of Engineering\\
      Tel Aviv University, 6997801 Tel Aviv, Israel
      \AND
      \name Tomer Michaeli \email tomer.m@ee.technion.ac.il\\
      \addr Faculty of Electrical and Computer Engineering\\
      Technion, 32000 Haifa, Israel}

\begin{document}

\maketitle

\begin{abstract}
Diffractive neural networks have recently emerged as a promising framework for all-optical computing. However, these networks are typically trained for a single task, limiting their potential adoption in systems requiring multiple functionalities. Existing approaches to achieving multi-task functionality either modify the mechanical configuration of the network per task or use a different illumination wavelength or polarization state for each task. 
In this work, we propose a new control mechanism, which is based on the illumination's angular spectrum. 
Specifically, we shape the illumination using an amplitude mask that selectively controls its angular spectrum. We employ different illumination masks for achieving different network functionalities, so that the mask serves as a unique task encoder. Interestingly, we show that effective control can be achieved over a very narrow angular range, within the paraxial regime. 
We numerically illustrate the proposed approach by training a single diffractive network to perform multiple image-to-image translation tasks. In particular, we demonstrate translating handwritten digits into typeset digits of different values, and translating handwritten English letters into typeset numbers and typeset Greek letters, where the type of the output is determined by the illumination's angular components.
As we show, the proposed framework can work under different coherence conditions, and can be combined with existing control strategies, such as different wavelengths. 
Our results establish the illumination angular spectrum as a powerful degree of freedom for controlling diffractive networks, enabling a scalable and versatile framework for multi-task all-optical computing.
\end{abstract}

\section{Introduction}
\label{introduction}

All-optical computing~\citep{shen2017deep, Lin2018, chang2018hybrid, wetzstein2020inference} has gained significant attention in recent years as it offers the potential for fast, efficient and parallel data processing~\citep{mcmahon2023physics}. In particular, all-optical diffractive neural networks~\citep{Lin2018} have emerged as a promising framework for free-space optical computing, which is especially suited for visual data. Diffractive networks comprise consecutive diffractive layers, usually in the form of phase masks, whose phases are optimized during a training stage to achieve some objective. 

Diffractive networks have proven a versatile framework capable of solving various tasks~\citep{Lin2018, qian2020performing, zhou2021large, bai2022image, chen2023photonic, bai2024information, chen2024superresolution, guo2025polarization, kleiner2025SR} under diverse illumination conditions~\citep{Luo2019broadband, kleiner2025coherence}. However, a single diffractive network is typically trained for solving a single task. This poses a significant limitation for systems requiring multiple functionalities, and stands in contrast with digital neural networks, which can incorporate various control mechanisms. For example, in deep image restoration, a single network can handle images degraded to varying degrees by incorporating a degradation map as an additional input~\citep{zhang2018learning, zhang2018ffdnet}. This map serves as a control signal to guide the network's functionality. 
 
To address this, recent work explored the potential of exploiting a single diffractive network for multiple tasks by using different control mechanisms. 
Some works used the physical configuration of the diffractive network as a control mechanism, for example, by removing, sharing or adding layers~\citep{li2021real, zhou2025automating, wang2025modular}, flipping layers~\citep{ren2025flippable} or rotating layers (\textit{e.g.}, by $180^{\circ}$)~\citep{li2023rubik, ma2024multiplexed, zhou2025automating}.  
Others controlled the operational mode by using different wavelengths~\citep{duan2023optical, li2023unidirectional, behroozinia2024leveraging} or polarization states~\citep{liu2025polarization, guo2025polarization, zhang2025integrated} for the light source illuminating the object at the network's input. The latter approach has been illustrated with metasurfaces~\citep{guo2025polarization, zhang2025integrated} and liquid crystals~\citep{liu2025polarization}. 
Many of these works focused on image classification, where different operational modes encode different datasets~\citep{li2021real, duan2023optical, li2023rubik, ren2025flippable, zhou2025automating}, so that the same network can classify \textit{e.g.}, either handwritten digits or fashion items according to the provided control signal. Other works utilized these control mechanisms for directional imaging~\citep{li2023unidirectional} and image encryption~\citep{wang2025modular, zhang2025integrated}. 

\begin{figure}
  \centering
   \includegraphics[width=\linewidth]{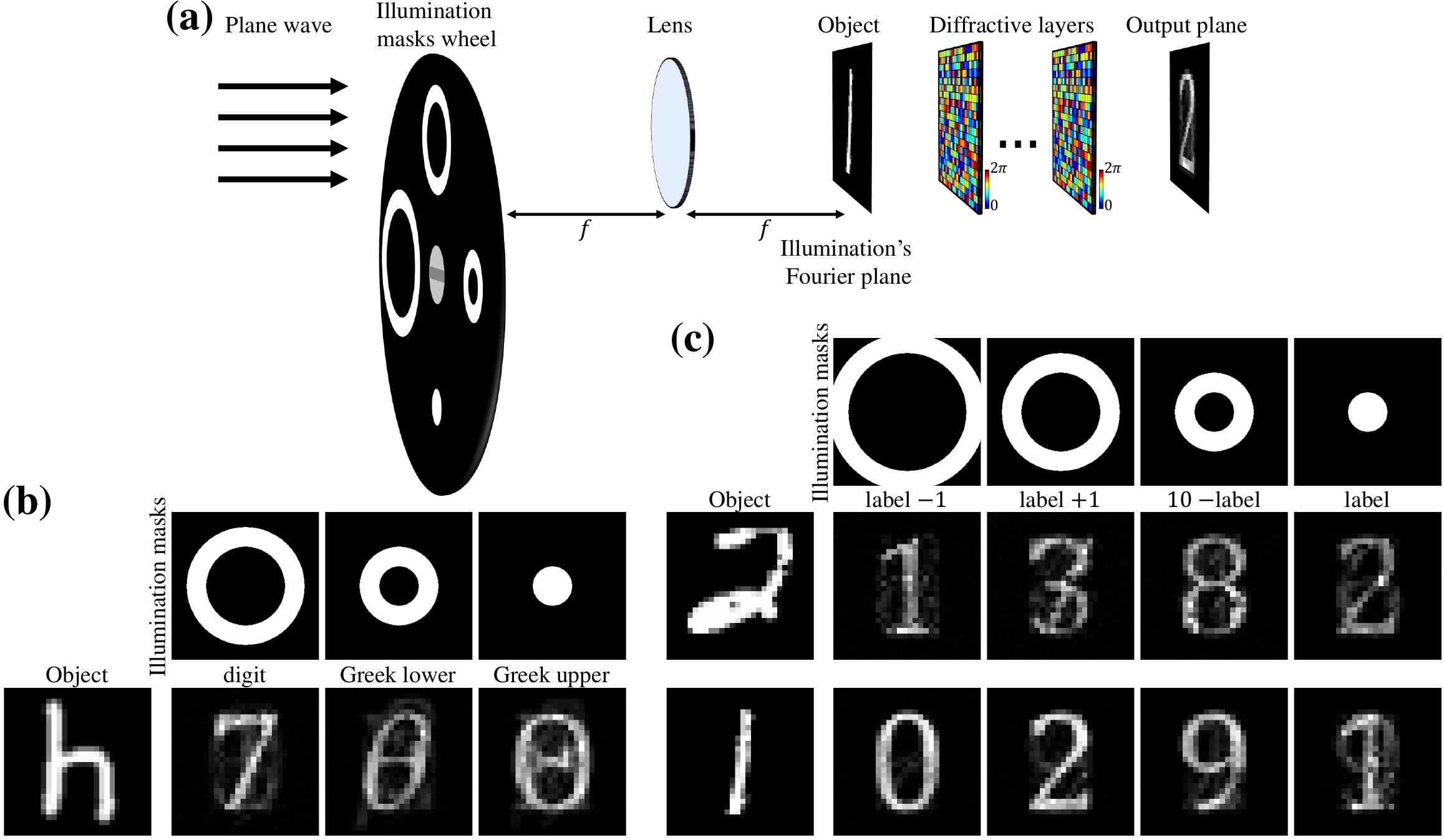}
   \caption{\textbf{Achieving multiple functionalities through illumination angular spectrum encoding.} (a) Schematic illustration of the proposed illumination angular spectrum encoding. An illumination mask is applied to a plane wave to retain only specific angular components. A $2f$ system maps the resulting field into its Fourier transform, creating an illumination profile that encodes a control signal for the diffractive network. The resulting profile then illuminates the object and propagates through the diffractive layers. Each illumination mask corresponds to a different task, enabling a single network to perform multiple functionalities. (b,c) Qualitative results for a single network trained with multiple predetermined ring-shaped illumination masks. In (b), a handwritten letter is transformed to either a digit or a lowercase Greek letter or an uppercase Greek letter, according to which one of three possible illumination masks is used. Similarly, in (c), a handwritten digit is transformed into one of four typeset digits depending on the value of the input digit and the mask used.}
   \label{fig:method}
\end{figure}

Here, we propose to utilize the angular spectrum of the light source illuminating the object as an additional degree of freedom for controlling the functionality of diffractive networks. 
We control the angular spectrum by passing a plane-wave illumination through an amplitude mask followed by a $2f$ system that maps the mask pattern to the angular spectrum at the object plane. 
The illuminated object is then processed by a diffractive network comprising a sequence of trainable phase masks. This encoding method operates within a narrow angular range and remains within the paraxial regime. 
By using different illumination masks, we train the same network to perform different tasks under different angular illumination conditions.
A schematic illustration of this process is shown in Fig.~\ref{fig:method}a.

Using the illumination direction as a control signal was explored in the context of reflective displays~\citep{Levin2013BRDF, Glasner2014reflectance}. There, the goal was to have the same display depict different images for different illumination directions. 
Using amplitude masks as part of a diffractive network was previously explored by Rahman et al.~\citep{rahman2021ensemble}, in a different setting and for a different purpose. Specifically, Rhaman et al.~trained multiple diffractive networks on the same classification task, each with a different optical preprocessing, including using amplitude masks. Their goal was to use an ensemble of these different networks to achieve higher classification accuracy than any single network, an approach commonly used in machine learning~\citep{ganaie2022ensemble}.   
In our work, we use masking directly on the illumination and not on the object, and we use it to enable the same network to perform multiple tasks rather than to train multiple networks on the same task. 
To the best of our knowledge, controlling the illumination angular spectrum was not utilized as a control mechanism for diffractive networks.

We numerically demonstrate the effectiveness of our proposed method on image-to-image translation tasks. 
These tasks involve transforming an input image into a corresponding output image and are ubiquitous in various domains, including biological and medical imaging~\citep{ronneberger2015u, rivenson2019virtual} and image restoration~\citep{zhang2018learning, zhang2018ffdnet}.
We illustrate three such tasks, one of translating handwritten digits to typeset digits of different values, one of translating handwritten English letters to either typeset digits or to typeset lowercase Greek letters or to uppercase Greek letters, and one of translating handwritten digits to typeset digits of different values, lowercase and uppercase Greek letters, and English letters. 
In all these tasks, the network's output is determined by the illumination's angular spectrum. 
Qualitative examples for the first two tasks are shown in Fig.~\ref{methods}b,c. 

It is important to note that a network trained for these image-to-image translation tasks needs to produce different outputs for the same input, when illuminating it by different patterns. This is in stark contrast with networks trained to classify multiple datasets, where images from different datasets exhibit different statistics. However, we also demonstrate the utility of our method for the simpler task of multi-dataset classification, which was studied by previous work~\citep{duan2023optical} using different control mechanisms.

We illustrate our approach with different kinds of illumination masks, including illumination masks that were jointly optimized with the diffractive network's phase masks. 
Additionally, we numerically illustrate the results of our method under broadband illumination, under spatially incoherent illumination~\citep{kleiner2025coherence}, and with wavelength encoding mechanisms~\citep{duan2023optical}, demonstrating the flexibility and generality of the proposed angular spectrum controlling mechanism.   

\section{Angular spectrum encoding}
\label{as-enc}

To control the functionality of a diffractive network, we propose illuminating the input object with light containing only specific angular components. 
We achieve this by passing a plane wave through an illumination mask, followed by a $2f$ system that preforms a Fourier transform. The resulting illumination pattern on the object plane exhibits angular components determined by the illumination mask, thereby enabling the desired control. 
The $2f$ system is simulated using a lens with a quadratic phase profile of dimensions $300 \times 300$ pixels and focal length of $f=20$ cm, with free-space propagation of $f$ before and after the lens. The lens' pixel dimensions are $10$ $\mu$m $\times 10$ $\mu$m, which results in an aperture radius of $r=1.5$ mm. This configuration translates into a maximal angle of $\theta_{\max}=\tan^{-1}(r/f)\approx0.43^\circ$, corresponding to a numerical aperture of $\mathrm{NA}=\sin(\theta_{\max})\approx7.5\times10^{-3}$. Namely, the proposed encoding method operates within a narrow angular range and fully within the paraxial regime. 
The illuminated object is followed by a diffractive network comprised of a sequence of trainable phase masks, each of dimensions $300 \times 300$ pixels, separated by $5$ cm of free-space propagation. The phase masks' pixel dimensions are again $10$ $\mu$m $\times 10$ $\mu$m.
Free-space propagation is simulated using the angular spectrum method and Rayleigh-Sommerfeld diffraction formulation~\citep{goodman2005introduction}, with appropriate zero-padding applied to all fields and optical elements to prevent aliasing. The following results are illustrated with four different diffractive layers. 
Unless specifically mentioned otherwise, all the following results are obtained under monochromatic illumination of 550 nm. Using multiple discrete wavelengths or broadband illumination is discussed in the next sections. Further implementation details can be found in Sec.~\ref{methods} and \ref{sm:forward}.  

The diffractive networks were trained using an $\ell_2$ regression loss between the each network's output, $\hat{I}$, and a set of predefined ground-truth images, $\{I_{\text{GT}}^{(c)}\}_{c=1}^C$, one for each task. For simplicity, we represent these quantities as column vectors, omitting their spatial coordinates. The appropriate ground-truth image was selected based on the task, which also determined the angular components of the light illuminating the input object, and based on the label of the input object. 
Specifically, the networks were trained by minimizing the loss
\begin{equation}\label{eq:loss}
    \mathcal{L} = \frac{1}{N} \sum_{i=1}^N \sum_{j=1}^M \left\|\hat{I}^{(i,j)} - I_{\text{GT}}^{(c)}\right\|^2,
\end{equation}
where $N$ is the number of training examples, $M$ is the number of illumination masks, and $c$ depends on both the input image label and the task. 
This training results in a network capable of mapping the same input image to different output images, depending on the angular components of the incident illumination. 

We demonstrate the following image-to-image translation setting. 
In the first setting, handwritten digits from the MNIST dataset~\citep{lecun1998mnist} are translated into one of ten typeset digits (0-9). We apply four different masks to the illumination source, each corresponding to a distinct transformation rule (a simple mathematical operation) applied to the input label: label plus one, label minus one, ten minus the label and the label itself. The input images are of labels 1-8 to avoid edge cases (see Sec.~\ref{methods} for additional details). A qualitative example is shown in Fig.~\ref{fig:method}c, where, for example, in the first row the input image is the handwritten digit `2' and the output images are the typeset digits `1',`3',`8' and `2'.      
In the second setting, the input objects are the first ten letters of the English alphabet, taken from the EMNIST dataset of handwritten letters~\citep{cohen2017emnist}. Here, we apply three different illumination masks to obtain at the network's output either images of typeset digits (0-9), or lowercase Greek letters, or uppercase Greek letters. The digits correspond to the alphabetical index of the English letters (starting from 0), while the Greek letters correspond to the equivalent position in the Greek alphabet. 
A qualitative example is shown in Fig.~\ref{fig:method}b, where the input image is the handwritten letter `h' and the network's outputs are `7', `$\theta$' and `$\Theta$', each corresponding to a different illumination mask. 
In the third setting, a single network is trained for eight different tasks, where a handwritten digit is translated into different digits, lowercase Greek, uppercase Greek, and English letters. Examples are shown in Fig.~\ref{fig:eight}.  The ground-truth typeset digits and letters are shown in \ref{sm:typeset}. Finally, to illustrate the generality of our approach we apply it to multi-dataset image classification in Sec.~\ref{sec:classification}, similarly to previous works~\citep{duan2023optical}.  We note that in all masks, white and black regions correspond to transparent and opaque areas, respectively. 

\subsection{Effect of different families of illumination masks}
\label{effect}

\begin{figure}
  \centering
   \includegraphics[width=\linewidth]{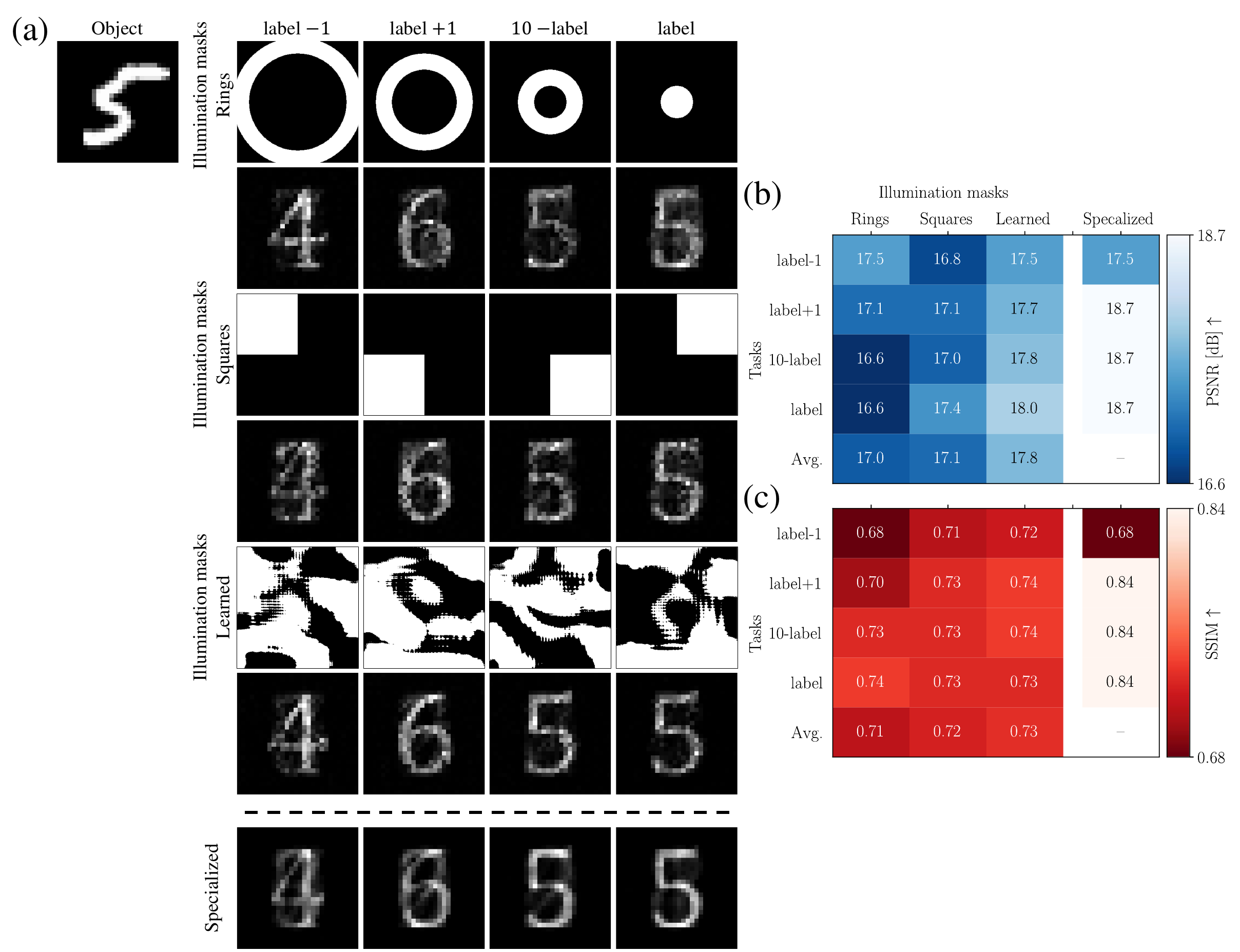}
   \caption{\textbf{Evaluation of angular-spectrum-encoded handwritten digit translation with different illumination masks.} (a) An input image of a handwritten `5' is translated into different typeset digits, with the target digit selected by the angular components of the incident illumination. Rows 1, 3, and 5 show the illumination masks (Rings, Squares, and Learned), while rows 2, 4, and 6 show examples of the corresponding networks' output. The seventh row, provided for comparison, shows outputs from four specialized networks, each trained using all illumination angular components but optimized to perform a single translation task. (b) PSNR and (c) SSIM averaged over the test set for the different networks. Performance of the specialized networks is shown in the rightmost column for reference.} 
   \label{fig:masks}
\end{figure}

The proposed angular spectrum encoding uses different illumination patterns to control the functionality of the diffractive network. We explore various mask designs and their effect on the network's performance. 
We quantified the similarity between the network's outputs and the ground-truth images using the peak signal-to-noise ratio (PSNR) and the structure similarity index (SSIM)~\citep{wang2004image}. For both metrics higher values indicate greater similarity. 
To ensure fair and consistent evaluation across different baselines, illumination mask shapes and configurations, we normalize the network's outputs. Specifically, we divide the intensity values at the network's output by their sum, taken over the entire field of view, and multiply the result by the sum of intensities of the corresponding ground-truth image.

We evaluate the performance of a network trained with the proposed angular spectrum encoding using two kinds of predefined illumination masks: ring-shaped and square-shaped. Both types of masks are binary, consisting of pixels that either fully pass or fully block the incident light. 
The ring illumination masks comprise a set of nonoverlapping cocentric rings, where the smallest one is a filled circle (a ring with inner radius zero), and the inner radius of each subsequent ring equals the outer radius of the previous one. 
The square illumination masks consist of single squares, each covering one quadrant of the mask.     
Both mask designs achieve comparable performances, quantitatively and qualitatively, as shown in Fig.~\ref{fig:masks} (indicated as ``Rings'' and ``Squares''). 

As an alternative to manually selecting the shapes of the illumination masks, we additionally explore jointly optimizing the masks along with the diffractive network's phase masks. In this approach, we constrain the illumination mask values to the range [0,1] by clamping their values to this range. 
We do not explicitly enforce binary masks during training to avoid difficulties in the optimization process. However, in practice, the learned masks tend to converge to nearly binary patterns, with only a few pixels taking values strictly between 0 and 1. Therefore, after training we threshold the patterns to obtain purely binary masks. This has a negligible effect on the results.  
Unlike the predefined masks, we do not restrict the learned masks to be non-overlapping. And indeed, the optimized masks have noticeable overlaps (Fig.~\ref{fig:masks}).
As illustrated in Fig.~\ref{fig:masks} in the context of the digit translation task and in Fig.~\ref{fig:masks_letters} in the context of the letter translation task, the learned illumination masks lead to better performance than the predefined ones. 
Specifically, the average PSNR gain achieved by using learned over predefined masks is  0.7 dB.  

For reference, we provide the results of specialized networks trained to perform only a single task. These networks, trained under monochromatic illumination of $550$ nm, achieve similar or higher PSNR and SSIM values compared to the multi-task networks trained with the proposed encoding method, as can be seen in the rightmost columns of Fig.~\ref{fig:masks}b,c. Qualitative results can be seen at the bottom row of Fig.~\ref{fig:masks}a. 

Similarly, we evaluated the effect of predefined ring-shaped illumination masks and learned illumination masks on the task of translating handwritten English letters to typeset digits and Greek letters. Again, the learned illumination masks yield better performance than the ring illumination masks. Qualitative and quantitative results are provided in \ref{sm:letters}. 

As mentioned above, the proposed encoding method operates within a narrow angular range. In \ref{sm:amp_phase}, we show the amplitude and phase at the object plane (the illumination's Fourier plane) resulting from the ring, square, and learned illumination masks, demonstrating the distinct profiles achieved within this narrow range.

\begin{figure}
  \centering
   \includegraphics[width=\linewidth]{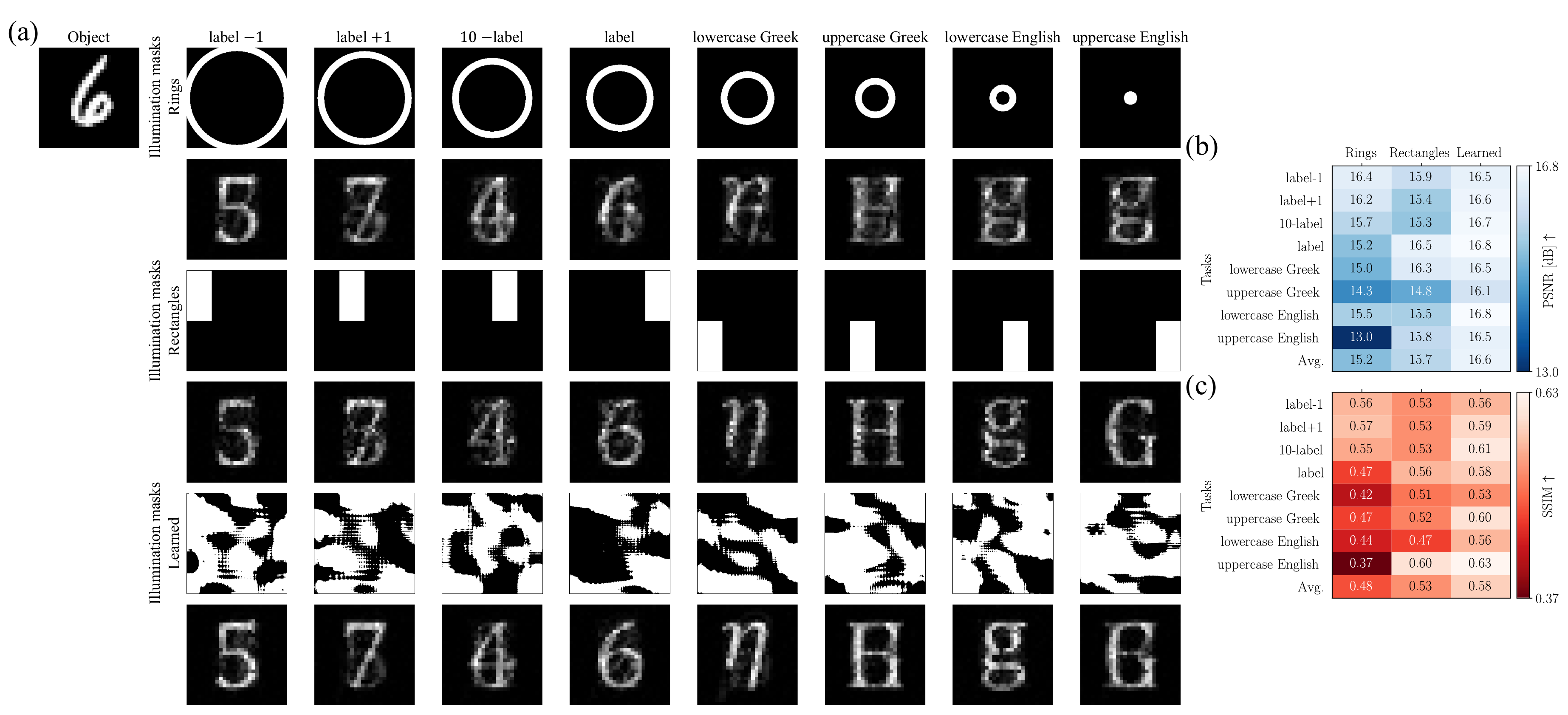}
   \caption{\textbf{Evaluation of angular-spectrum-encoded handwritten digit translation with different illumination masks for eight tasks.} (a) An input image of a handwritten `6' is translated into different typeset digits, Greek and English letters by a single network. The target output is determined by the angular components of the illumination. Rows 1, 3, and 5 show the illumination masks (Rings, Rectangles and Learned), while rows 2, 4, and 6 show examples of the corresponding networks' output. (b) PSNR and (c) SSIM averaged over the test set for the different networks.} 
   \label{fig:eight}
\end{figure}

We further demonstrate the proposed method's performance for simultaneously solving eight image-to-image tasks. In this setting, the input image is a handwritten digit and the output images include: label minus one, label plus one, ten minus the label, the label itself, and lowercase and uppercase letters from both the Greek and English alphabets. The letters correspond to the alphabetical index of the input digit (starting from 0). 
Figure \ref{fig:eight} shows the results obtained with predefined ring-shaped and rectangle-shaped masks, as well as with learned masks. The rectangle-shaped masks are similar to the square-shaped illustrated in Fig.~\ref{fig:masks}, where each cover half of the mask height and a quarter of its width.
In this setting, the learned masks significantly outperform the predefined masks, achieving an average PSNR gain of 1.4 dB over the ring-shaped masks and 0.9 dB over the rectangle-shaped ones. In particular, the ring-shaped masks occasionally fail on certain tasks, for example, producing a lowercase English letter instead of an uppercase English letter in the top two rows. 
These results suggest that while predefined masks can handle a small number of tasks to a reasonable extent, their performance can significantly deteriorate as the number of tasks increases, if not chosen properly. In such cases, the advantage of learned masks becomes more pronounced.

\subsection{Combining angular spectrum encoding with wavelength encoding}

\begin{figure}
  \centering
   \includegraphics[width=\linewidth]{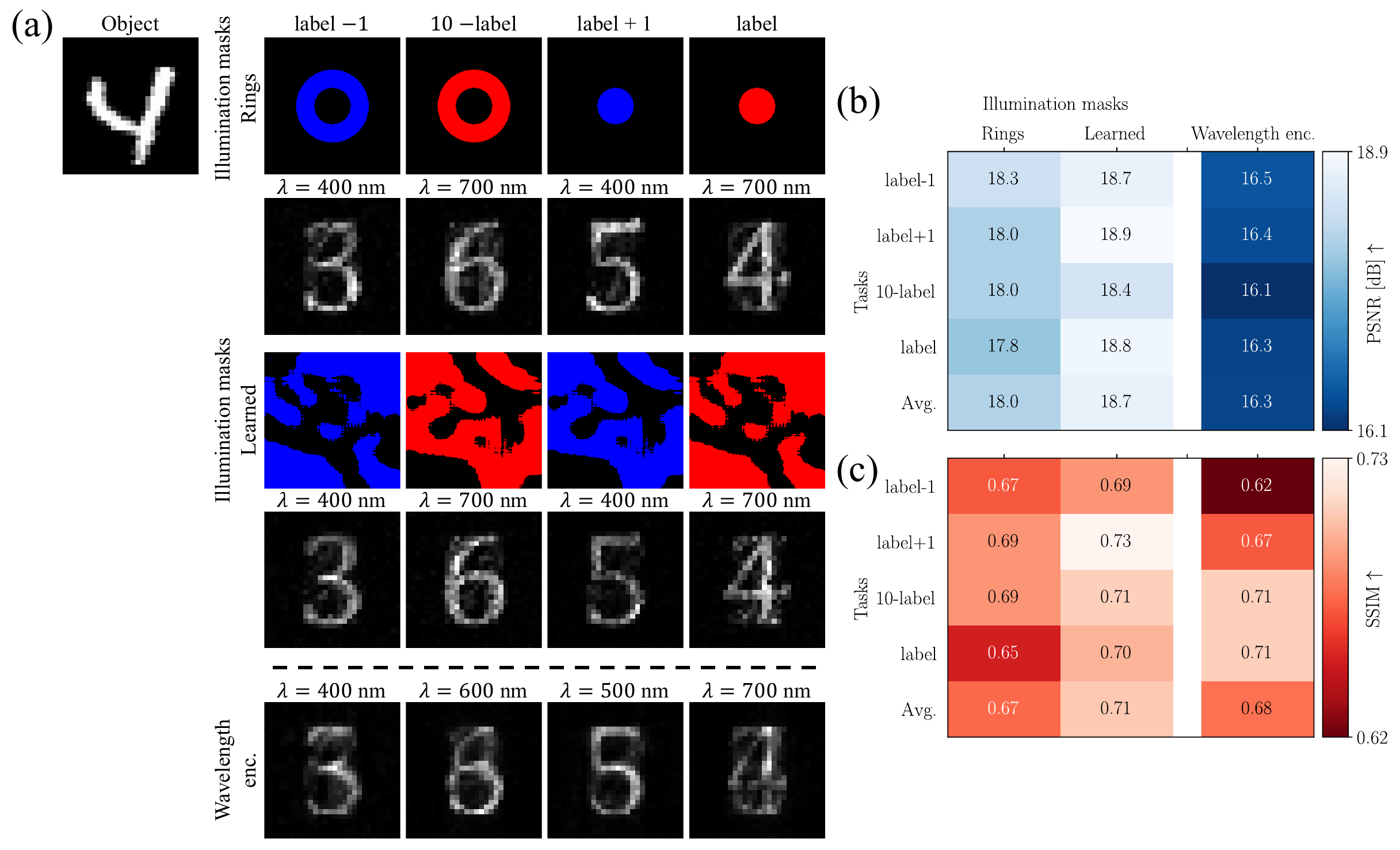}
   \caption{\textbf{Combining angular spectrum encoding and wavelength encoding.} (a) An input image of a handwritten `4' is translated into different typeset digits. Rows 1 and 3 show different illumination masks (Rings and Learned), while rows 2 and 4 show the corresponding outputs of a single network trained with each mask type and a distinct wavelength, indicated by the color of the illumination mask (400 or 700 nm, detailed above each image). The fifth row illustrates the results of a network trained with wavelength encoding, the used wavelength for each task is indicated above each image. (b) PSNR and (c) SSIM averaged over the test set for the different networks.}
   \label{fig:wavelengths}
\end{figure}

A common approach for controlling the functionality of a diffractive network is to vary the wavelength of the monochromatic illumination~\citep{duan2023optical, li2023unidirectional, behroozinia2024leveraging}. We refer to this approach as wavelength encoding. To evaluate this approach against the proposed angular spectrum encoding, we trained a diffractive network under monochromatic illumination with distinct wavelengths assigned to each task. Specifically, we trained the network to translate handwritten digits into typeset digits of different values, as described above, using wavelengths of 400, 500, 600 and 700 nm to encode the four different tasks. The same hyperparameters and training procedure were used as in the angular spectrum encoding numerical experiments.  
The network trained with wavelength encoding performs worse than the network trained with angular spectrum encoding, as shown qualitatively and quantitatively at the bottom of Fig.~\ref{fig:wavelengths}, achieving $\sim$2 dB lower average PSNR compared to networks trained with the proposed method.

We further combined angular spectrum encoding with wavelength encoding by training a network using a set of distinct wavelengths and illumination masks, where each unique combination encodes a different task. For example, in the handwritten-to-typeset digit translation task, we trained a diffractive network with two wavelengths (400 and 700 nm) and two illumination masks, enabling four distinct tasks. We trained such networks using both predefined ring-shaped illumination masks and learned illumination masks jointly optimized with the diffractive network's phase masks. 
Combining both encoding methods yields improved performance across all tasks compared to either encoding method alone, as illustrated in Fig.~\ref{fig:wavelengths}.
We additionally trained a network with two wavelengths and four different illumination masks, enabling eight distinct tasks, as in Fig.~\ref{fig:eight}. Results achieved by this network are shown in \ref{sm:eight}.

\subsection{Broadband illumination and spatially incoherent illumination}

Diffractive networks operating under broadband illumination or spatially incoherent illumination have gained significant attention in recent years~\citep{Luo2019broadband, kleiner2025coherence, jia2024partially, qin2025all, rahman2023universal}, as these illumination conditions better reflect natural environments. 
We next illustrate that the proposed angular spectrum encoding method is easily adaptable to such conditions. 

\begin{figure}
  \centering
   \includegraphics[width=0.4\linewidth]{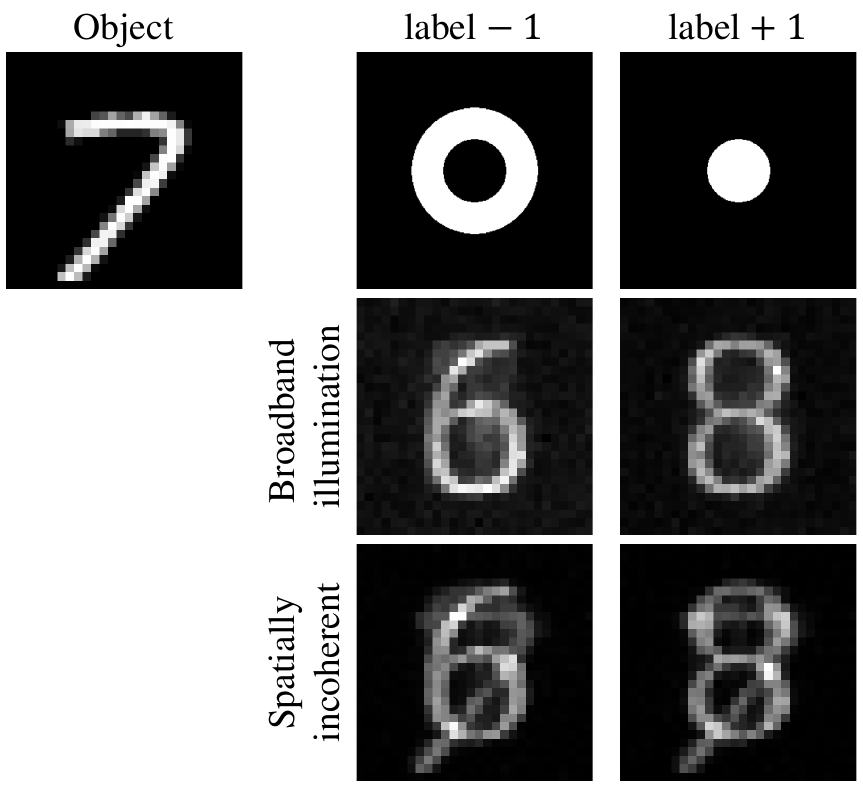}
   \caption{\textbf{Angular spectrum encoding with different illumination conditions.} 
   An input image of a handwritten `7' is translated into different typeset digits.
   The second row illustrates the results of a network trained with spatially coherent broadband illumination. Note the speckle patterns visible in the background. The third row illustrates the results of a network trained with spatially incoherent monochromatic illumination.}
   \label{fig:other_illum}
\end{figure}

We trained diffractive networks under spatially coherent broadband illumination, and under spatially incoherent monochromatic illumination. Both networks were trained to translate handwritten digits into two typeset digits corresponding to the input image label minus one and plus one, using ring-shaped illumination masks. Qualitative results for both networks are shown in Fig.~\ref{fig:other_illum}.  

For the spatially coherent broadband illumination, we trained a network using 32 wavelengths, uniformly sampled between 400 -- 700 nm, serving as a discrete approximation of a continuous spectrum. For evaluation, we used 500 wavelengths randomly sampled within the same range.  
This network achieves average PSNR values of 14.3 dB and 14.5 dB for producing the input image label minus one and plus one, respectively. The output images produced by this network exhibit noticeable speckle patterns in the background (Fig.~\ref{fig:other_illum}), which contribute to the relatively low PSNR. These speckle patterns can be partially overcome by using a partially coherent illumination source~\citep{peng2021speckle, kleiner2025coherence}.   

For the spatially incoherent illumination, we used modal expansion~\citep{wolf1982new, rahman2023universal}. Namely, we multiplied the illumination source by a random phase pattern, $\exp\{i\phi(x,y)\}$, where $\{\phi(x,y)\}$ are independent random variables drawn from the uniform distribution $\mathcal{U}[0,2\pi]$, and coherently propagated the resulting field through the entire system up to the output plane, where we recorded the intensity. By repeating this process $L$ times and averaging the intensity results, we approximated incoherent propagation. This approximation becomes exact as $L \rightarrow \infty$. 
We trained the network with $L=25$ repetitions and evaluated it with $L=10000$ repetitions to better approximate the spatially incoherent illumination. 
This network achieves average PSNR values of 16.1 dB and 16.7 dB for producing the input image label minus one and plus one, respectively.

We chose the above settings for each network to fit our memory and computation requirements and to keep training time manageable.

\subsection{Angular spectrum encoding for multi-dataset classification}
\label{sec:classification}

Although the primary focus of this work is on image-to-image translation, we also validate the applicability of our approach to multi-dataset image classification. Specifically, we used angular spectrum encoding to train a single diffractive network capable of classifying images from the MNIST~\citep{lecun1998mnist}, FashionMNIST~\citep{xiao2017fashion}, KMNIST~\citep{clanuwat2018KMNIST} and EMNIST~\citep{cohen2017emnist} datasets, covering handwritten digits, fashion items, handwritten Japanese letters and handwritten English letters, repetitively. The network is informed of the input type through the illumination pattern.  
Thus, the network's input changes across datasets, and the unique illumination pattern implicitly encodes these statistical differences. This is fundamentally different from image-to-image translation (as described above), where the same input image is mapped to different output images, depending on the illumination's angular components. 
We compared networks trained with angular spectrum encoding, wavelength encoding~\citep{duan2023optical} and without encoding. The network trained with angular spectrum encoding achieved the highest average classification accuracy among the three. Additional details are provided in \ref{sm:classification}.
 
\section{Conclusion}
\label{conclusion}

We demonstrated the effectiveness and generality of angular spectrum encoding for controlling the functionality of diffractive networks and enabling multi-task performance. 
This encoding method is general, applicable under various illumination conditions and can be combined with other encoding methods, such as wavelength encoding. 
The proposed approach, which fully operates in the paraxial regime, introduces a previously unexplored degree of freedom in designing and controlling diffractive networks, which opens new opportunities for scalable, multi-functional all-optical computing across diverse applications.

\section{Materials and methods}
\label{methods}

Training our diffractive networks requires a dataset $\mathcal{D}$ of objects (images), $\mathcal{D}=\{I^{(i)}\}_{i=1}^N$. 
We used the MNIST \citep{lecun1998mnist} and EMNIST \citep{cohen2017emnist}
datasets, which contain $28\times 28$ grayscale images of handwritten digits and letters, respectively. The set of ground-truth images, $\{I_{\text{GT}}^{(c)}\}_{c=1}^C$ was manually created and processed. The ground-truth images are illustrated in \ref{sm:typeset}.  

Our networks were trained using a regression loss between the network's output and the ground-truth images (Sec.~\ref{as-enc}). While this training yielded the results described above, it came at the expense of low energy preservation along the optical path, where only $\sim$8\%-1\% of the input energy was preserved, depending on the specific configuration. 
This phenomenon can be mitigated to some extent by replacing the loss of Eq.~\eqref{eq:loss} with a regression loss between the network's normalized output and the ground-truth image and adding an additional loss term that penalizes low output intensities. By tuning a hyperparameter that balances these two loss terms, we can traverse the tradeoff between output fidelity and energy preservation. Details on this approach are provided in \ref{sm:energy}.

For translating handwritten digits to typeset digits, we used input images from the MNIST dataset with labels 1-8 to avoid edge cases involving `9' or `0'. We used $\sim1000$ images per class, resulting in a training set of $8000$ images. 
Evaluation was performed on the entire MNIST test set, restricted to labels 1–8, yielding $8000$ test images. 
For translating handwritten English letters to typeset digits or Greek letters, we used the first ten letters from EMNIST, with $5000$ training examples. Evaluation was performed on the full EMNIST test set, again restricted to the first ten letters.

We trained all networks using the Adam optimizer~\citep{kingma2015adam} with a learning rate of $\eta=10^{-2}$ and a scheduler that reduces the learning rate by a factor of $2$ every $100$ epochs. 
Simulations were implemented using the Pytorch deep learning framework~\citep{paszke2019pytorch} and were executed on a Linux machine with NVIDIA GeForce GTX 2080 Ti GPU. 
For spatially coherent and monochromatic illumination, we trained the network for $1000$ epochs, resulting in $\sim 3$ training hours. 
For broadband and spatially incoherent illumination, we trained for only $250$ epochs to keep training time reasonable, resulting in $\sim6$ hours for spatially incoherent illumination and $\sim20$ hours for broadband illumination.  

Details regarding the forward propagation, both for coherent and for incoherent illumination, are provided in \ref{sm:forward}.

\medskip
\textbf{Data Availability Statement}
The data that support the findings of this study is available at \cite{github}. 
This data was derived from the following resources available in the public domain: 
\begin{itemize}
    \item MNIST dataset - \url{docs.pytorch.org/vision/main/generated/torchvision.datasets.MNIST}
    \item EMNIST dataset - \url{https://www.nist.gov/itl/products-and-services/emnist-dataset}
    \item FashionMNIST dataset - \url{https://github.com/zalandoresearch/fashion-mnist}
    \item KMNIST dataset - \url{https://github.com/rois-codh/kmnist}
\end{itemize}

\clearpage

\bibliography{main}

@article{Lin2018,
   author = {Xing Lin and Yair Rivenson and Nezih T. Yardimci and Muhammed Veli and Yi Luo and Mona Jarrahi and Aydogan Ozcan},
   doi = {10.1126/science.aat8084},
   issn = {10959203},
   issue = {6406},
   journal = {Science},
   month = {9},
   pages = {1004-1008},
   pmid = {30049787},
   publisher = {American Association for the Advancement of Science},
   title = {All-optical machine learning using diffractive deep neural networks},
   volume = {361},
   year = {2018},
}

@article{Luo2019broadband,
   author = {Yi Luo and Deniz Mengu and Nezih T. Yardimci and Yair Rivenson and Muhammed Veli and Mona Jarrahi and Aydogan Ozcan},
   doi = {10.1038/s41377-019-0223-1},
   issn = {20477538},
   issue = {1},
   journal = {Light: Science and Applications},
   month = {12},
   publisher = {Springer Nature},
   title = {Design of task-specific optical systems using broadband diffractive neural networks},
   volume = {8},
   year = {2019},
}

@article{rahman2021ensemble,
  title={Ensemble learning of diffractive optical networks},
  author={Rahman, Md Sadman Sakib and Li, Jingxi and Mengu, Deniz and Rivenson, Yair and Ozcan, Aydogan},
  journal={Light: Science \& Applications},
  volume={10},
  number={1},
  pages={1--13},
  year={2021},
  publisher={Nature Publishing Group}
}

@article{wetzstein2020inference,
  title={Inference in artificial intelligence with deep optics and photonics},
  author={Wetzstein, Gordon and Ozcan, Aydogan and Gigan, Sylvain and Fan, Shanhui and Englund, Dirk and Solja{\v{c}}i{\'c}, Marin and Denz, Cornelia and Miller, David AB and Psaltis, Demetri},
  journal={Nature},
  volume={588},
  number={7836},
  pages={39--47},
  year={2020},
  publisher={Nature Publishing Group}
}

@article{chang2018hybrid,
  title={Hybrid optical-electronic convolutional neural networks with optimized diffractive optics for image classification},
  author={Chang, Julie and Sitzmann, Vincent and Dun, Xiong and Heidrich, Wolfgang and Wetzstein, Gordon},
  journal={Scientific reports},
  volume={8},
  number={1},
  pages={1--10},
  year={2018},
  publisher={Nature Publishing Group}
}

@article{kingma2015adam,
  title={Adam: A method for stochastic optimization},
  author={Kingma, Diederik P and Ba, Jimmy},
  journal={International Conference on Learning Representations},
  volume={3},
  year={2015}
}

@article{chen2023photonic,
  title={Photonic unsupervised learning variational autoencoder for high-throughput and low-latency image transmission},
  author={Chen, Yitong and Zhou, Tiankuang and Wu, Jiamin and Qiao, Hui and Lin, Xing and Fang, Lu and Dai, Qionghai},
  journal={Science Advances},
  volume={9},
  number={7},
  pages={eadf8437},
  year={2023},
  publisher={American Association for the Advancement of Science}
}

@article{bai2022image,
  title={To image, or not to image: class-specific diffractive cameras with all-optical erasure of undesired objects},
  author={Bai, Bijie and Luo, Yi and Gan, Tianyi and Hu, Jingtian and Li, Yuhang and Zhao, Yifan and Mengu, Deniz and Jarrahi, Mona and Ozcan, Aydogan},
  journal={eLight},
  volume={2},
  number={1},
  pages={1--20},
  year={2022},
  publisher={SpringerOpen}
}

@article{duan2023optical,
  title={Optical multi-task learning using multi-wavelength diffractive deep neural networks},
  author={Duan, Zhengyang and Chen, Hang and Lin, Xing},
  journal={Nanophotonics},
  year={2023},
  publisher={De Gruyter}
}

@book{goodman2005introduction,
  title={Introduction to Fourier optics},
  author={Goodman, Joseph W},
  year={2005},
  pages={55--61},
  publisher={Roberts and Company publishers},
}

@article{paszke2019pytorch,
  title={Pytorch: An imperative style, high-performance deep learning library},
  author={Paszke, Adam and Gross, Sam and Massa, Francisco and Lerer, Adam and Bradbury, James and Chanan, Gregory and Killeen, Trevor and Lin, Zeming and Gimelshein, Natalia and Antiga, Luca and others},
  journal={Advances in neural information processing systems},
  volume={32},
  year={2019}
}

@article{shen2017deep,
  title={Deep learning with coherent nanophotonic circuits},
  author={Shen, Yichen and Harris, Nicholas C and Skirlo, Scott and Prabhu, Mihika and Baehr-Jones, Tom and Hochberg, Michael and Sun, Xin and Zhao, Shijie and Larochelle, Hugo and Englund, Dirk and others},
  journal={Nature photonics},
  volume={11},
  number={7},
  pages={441--446},
  year={2017},
  publisher={Nature Publishing Group UK London}
}

@article{li2021real,
  title={Real-time multi-task diffractive deep neural networks via hardware-software co-design},
  author={Li, Yingjie and Chen, Ruiyang and Sensale-Rodriguez, Berardi and Gao, Weilu and Yu, Cunxi},
  journal={Scientific Reports},
  volume={11},
  number={1},
  pages={11013},
  year={2021},
  publisher={Nature Publishing Group UK London}
}

@article{zhou2021large,
  title={Large-scale neuromorphic optoelectronic computing with a reconfigurable diffractive processing unit},
  author={Zhou, Tiankuang and Lin, Xing and Wu, Jiamin and Chen, Yitong and Xie, Hao and Li, Yipeng and Fan, Jingtao and Wu, Huaqiang and Fang, Lu and Dai, Qionghai},
  journal={Nature Photonics},
  volume={15},
  number={5},
  pages={367--373},
  year={2021},
  publisher={Nature Publishing Group UK London}
}

@article{rahman2023universal,
  title={Universal linear intensity transformations using spatially incoherent diffractive processors},
  author={Rahman, Md Sadman Sakib and Yang, Xilin and Li, Jingxi and Bai, Bijie and Ozcan, Aydogan},
  journal={Light: Science \& Applications},
  volume={12},
  number={1},
  pages={195},
  year={2023},
  publisher={Nature Publishing Group UK London}
}

@article{qian2020performing,
  title={Performing optical logic operations by a diffractive neural network},
  author={Qian, Chao and Lin, Xiao and Lin, Xiaobin and Xu, Jian and Sun, Yang and Li, Erping and Zhang, Baile and Chen, Hongsheng},
  journal={Light: Science \& Applications},
  volume={9},
  number={1},
  pages={59},
  year={2020},
  publisher={Nature Publishing Group UK London}
}

@article{lecun1998mnist,
  title={The MNIST database of handwritten digits},
  author={LeCun, Yann},
  journal={http://yann. lecun. com/exdb/mnist/},
  year={1998}
}

@article{xiao2017fashion,
  title={Fashion-mnist: a novel image dataset for benchmarking machine learning algorithms},
  author={Xiao, Han and Rasul, Kashif and Vollgraf, Roland},
  journal={arXiv preprint arXiv:1708.07747},
  year={2017}
}

@article{guo2025polarization,
  title={Polarization-selective unidirectional and bidirectional diffractive neural networks for information security and sharing},
  author={Guo, Ziqing and Tan, Zhiyu and Zang, Xiaofei and Zhang, Teng and Wang, Guannan and Li, Hongguang and Wang, Yuanbo and Zhu, Yiming and Ding, Fei and Zhuang, Songlin},
  journal={Nature Communications},
  volume={16},
  number={1},
  pages={4492},
  year={2025},
  publisher={Nature Publishing Group UK London}
}

@article{li2023unidirectional,
  title={Unidirectional imaging using deep learning--designed materials},
  author={Li, Jingxi and Gan, Tianyi and Zhao, Yifan and Bai, Bijie and Shen, Che-Yung and Sun, Songyu and Jarrahi, Mona and Ozcan, Aydogan},
  journal={Science Advances},
  volume={9},
  number={17},
  pages={eadg1505},
  year={2023},
  publisher={American Association for the Advancement of Science}
}

@inproceedings{li2023rubik,
  title     = {Rubik's Optical Neural Networks: Multi-task Learning with Physics-aware Rotation Architecture},
  author    = {Li, Yingjie and Gao, Weilu and Yu, Cunxi},
  booktitle = {Proceedings of the Thirty-Second International Joint Conference on
               Artificial Intelligence, {IJCAI-23}},
  publisher = {International Joint Conferences on Artificial Intelligence Organization},
  pages     = {7197--7206},
  year      = {2023},
  month     = {8},
  doi       = {10.24963/ijcai.2023/847},
}

@article{behroozinia2024leveraging,
  title={Leveraging multiplexed metasurfaces for multi-task learning with all-optical diffractive processors},
  author={Behroozinia, Sahar and Gu, Qing},
  journal={Nanophotonics},
  volume={13},
  number={24},
  pages={4505--4517},
  year={2024},
  publisher={De Gruyter}
}

@article{zhou2025automating,
  title={Automating multi-task learning on optical neural networks with weight sharing and physical rotation},
  author={Zhou, Shanglin and Li, Yingjie and Gao, Weilu and Yu, Cunxi and Ding, Caiwen},
  journal={Scientific Reports},
  volume={15},
  number={1},
  pages={14419},
  year={2025},
  publisher={Nature Publishing Group UK London}
}

@article{bai2024information,
  title={Information-hiding cameras: Optical concealment of object information into ordinary images},
  author={Bai, Bijie and Lee, Ryan and Li, Yuhang and Gan, Tianyi and Wang, Yuntian and Jarrahi, Mona and Ozcan, Aydogan},
  journal={Science Advances},
  volume={10},
  number={24},
  pages={eadn9420},
  year={2024},
  publisher={American Association for the Advancement of Science}
}

@article{mcmahon2023physics,
  title={The physics of optical computing},
  author={McMahon, Peter L},
  journal={Nature Reviews Physics},
  volume={5},
  number={12},
  pages={717--734},
  year={2023},
  publisher={Nature Publishing Group UK London}
}

@article{kleiner2025coherence,
author = {Kleiner, Matan and Michaeli, Lior and Michaeli, Tomer},
title = {Coherence Awareness in Diffractive Neural Networks},
journal = {Laser \& Photonics Reviews},
volume = {19},
number = {10},
pages = {2401299},
year = {2025}
}

@article{jia2024partially,
  title={Partially coherent diffractive optical neural network},
  author={Jia, Qi and Shi, Bojian and Zhang, Yanxia and Li, Hang and Li, Xiaoxin and Feng, Rui and Sun, Fangkui and Cao, Yongyin and Wang, Jian and Qiu, Cheng-Wei and others},
  journal={Optica},
  volume={11},
  number={12},
  pages={1742--1749},
  year={2024},
  publisher={Optica Publishing Group}
}

@article{clanuwat2018KMNIST,
    title	= {Deep Learning for Classical Japanese Literature},
    author={Clanuwat, Tarin and Bober-Irizar, Mikel and Kitamoto, Asanobu and Lamb, Alex and Yamamoto, Kazuaki and Ha, David},
    year	= {2018},
    journal	= {Machine Learning for Creativity and Design Workshop in Advances in Neural Information Processing Systems}}

@inproceedings{cohen2017emnist,
  title={EMNIST: Extending MNIST to handwritten letters},
  author={Cohen, Gregory and Afshar, Saeed and Tapson, Jonathan and Van Schaik, Andre},
  booktitle={2017 international joint conference on neural networks (IJCNN)},
  pages={2921--2926},
  year={2017},
  organization={IEEE}
}

@article{liu2025polarization,
author = {Mengqin Liu and Xianglin Ye and Yingjie Zhou and Dongliang Tang and Fan Fan},
journal={Optics Letters},
number = {17},
pages = {5446--5449},
publisher = {Optica Publishing Group},
title = {Polarization-multiplexed diffractive neural networks for multi-task classification based on liquid crystals},
volume = {50},
month = {Sep},
year = {2025},
doi = {10.1364/OL.571548},
}

@article{Glasner2014reflectance,
author = {Glasner, Daniel and Zickler, Todd and Levin, Anat},
title = {A reflectance display},
year = {2014},
issue_date = {July 2014},
publisher = {Association for Computing Machinery},
address = {New York, NY, USA},
volume = {33},
number = {4},
issn = {0730-0301},
doi = {10.1145/2601097.2601140},
journal = {ACM Trans. Graph.},
month = jul,
articleno = {61},
numpages = {12},
}

@article{Levin2013BRDF,
author = {Levin, Anat and Glasner, Daniel and Xiong, Ying and Durand, Fr\'{e}do and Freeman, William and Matusik, Wojciech and Zickler, Todd},
title = {Fabricating BRDFs at high spatial resolution using wave optics},
year = {2013},
issue_date = {July 2013},
publisher = {Association for Computing Machinery},
address = {New York, NY, USA},
volume = {32},
number = {4},
issn = {0730-0301},
doi = {10.1145/2461912.2461981},
journal = {ACM Trans. Graph.},
month = jul,
articleno = {144},
numpages = {14},
}

@article{kleiner2025SR,
url = {https://doi.org/10.1515/nanoph-2025-0294},
title = {Can the success of digital super-resolution networks be transferred to passive all-optical systems?},
author = {Matan Kleiner and Lior Michaeli and Tomer Michaeli},
pages = {3181--3190},
volume = {14},
number = {19},
journal = {Nanophotonics},
doi = {doi:10.1515/nanoph-2025-0294},
year = {2025},
lastchecked = {2025-11-17}
}

@article{ma2024multiplexed,
  title={Multiplexed All-Optical Permutation Operations Using a Reconfigurable Diffractive Optical Network},
  author={Ma, Guangdong and Yang, Xilin and Bai, Bijie and Li, Jingxi and Li, Yuhang and Gan, Tianyi and Shen, Che-Yung and Zhang, Yijie and Li, Yuzhu and I{\c{s}}{\i}l, {\c{C}}a{\u{g}}atay and others},
  journal={Laser \& Photonics Reviews},
  volume={18},
  number={11},
  pages={2400238},
  year={2024},
  publisher={Wiley Online Library}
}

@article{ren2025flippable,
  title={Flippable multitask diffractive neural networks based on double-sided metasurfaces},
  author={Ren, He and Zhou, Shuai and Feng, Yuxiang and Wang, Di and Yang, Xu and Chen, Shouqian},
  journal={Optics Letters},
  volume={50},
  number={6},
  pages={1997--2000},
  year={2025},
  publisher={Optica Publishing Group}
}

@article{wang2025modular,
  title={Modular Diffractive Neural Networks Using Cascaded Metasurfaces},
  author={Wang, Guannan and Zang, Xiaofei and Tan, Zhiyu and Zhang, Teng and Gao, Zhe and Wang, Yuanbo and Zhang, Deng and Shkurinov, Alexander P and Zhu, Yiming and Zhuang, Songlin},
  journal={Laser \& Photonics Reviews},
  volume={19},
  number={22},
  pages={e00923},
  year={2025},
  publisher={Wiley Online Library}
}

@article{zhang2025integrated,
  title={Integrated Polarization, Distance, and Rotation for Multi-DoF Diffractive Processor and Information Encryption},
  author={Zhang, Teng and Zang, Xiaofei and Tan, Zhiyu and Wang, Guannan and Guo, Ziqing and Gao, Zhe and Shkurinov, Alexander P and Ding, Fei and Zhu, Yiming and Zhuang, Songlin},
  journal={Advanced Materials},
  pages={2506222},
  year={2025},
  publisher={Wiley Online Library}
}

@inproceedings{ronneberger2015u,
  title={U-net: Convolutional networks for biomedical image segmentation},
  author={Ronneberger, Olaf and Fischer, Philipp and Brox, Thomas},
  booktitle={International Conference on Medical image computing and computer-assisted intervention},
  pages={234--241},
  year={2015},
  organization={Springer}
}

@article{rivenson2019virtual,
  title={Virtual histological staining of unlabelled tissue-autofluorescence images via deep learning},
  author={Rivenson, Yair and Wang, Hongda and Wei, Zhensong and de Haan, Kevin and Zhang, Yibo and Wu, Yichen and G{\"u}nayd{\i}n, Harun and Zuckerman, Jonathan E and Chong, Thomas and Sisk, Anthony E and others},
  journal={Nature biomedical engineering},
  volume={3},
  number={6},
  pages={466--477},
  year={2019},
  publisher={Nature Publishing Group UK London}
}

@article{wang2004image,
  title={Image quality assessment: from error visibility to structural similarity},
  author={Wang, Zhou and Bovik, Alan C and Sheikh, Hamid R and Simoncelli, Eero P},
  journal={IEEE transactions on image processing},
  volume={13},
  number={4},
  pages={600--612},
  year={2004},
  publisher={IEEE}
}

@article{qin2025all,
  title={All-optical Fourier neural network using partially coherent light},
  author={Qin, Jianwei and Liu, Yanbing and Liu, Yan and Liu, Xun and Li, Wei and Ye, Fangwei},
  journal={Chip},
  pages={100140},
  year={2025},
  publisher={Elsevier}
}

@article{wolf1982new,
  title={New theory of partial coherence in the space--frequency domain. Part I: spectra and cross spectra of steady-state sources},
  author={Wolf, Emil},
  journal={JOSA},
  volume={72},
  number={3},
  pages={343--351},
  year={1982},
  publisher={Optica Publishing Group}
}

@article{peng2021speckle,
  title={Speckle-free holography with partially coherent light sources and camera-in-the-loop calibration},
  author={Peng, Yifan and Choi, Suyeon and Kim, Jonghyun and Wetzstein, Gordon},
  journal={Science advances},
  volume={7},
  number={46},
  pages={eabg5040},
  year={2021},
  publisher={American Association for the Advancement of Science}
}

@article{chen2024superresolution,
  title={Superresolution imaging using superoscillatory diffractive neural networks},
  author={Chen, Hang and Gao, Sheng and Zhang, Haiou and Zhao, Zejia and Duan, Zhengyang and Wetzstein, Gordon and Lin, Xing},
  journal={Advanced Photonics},
  volume={6},
  number={5},
  pages={056004--056004},
  year={2024},
  publisher={Society of Photo-Optical Instrumentation Engineers}
}

@article{ganaie2022ensemble,
  title={Ensemble deep learning: A review},
  author={Ganaie, Mudasir A and Hu, Minghui and Malik, Ashwani Kumar and Tanveer, Muhammad and Suganthan, Ponnuthurai N},
  journal={Engineering Applications of Artificial Intelligence},
  volume={115},
  pages={105151},
  year={2022},
  publisher={Elsevier}
}

@inproceedings{zhang2018learning,
  title={Learning a single convolutional super-resolution network for multiple degradations},
  author={Zhang, Kai and Zuo, Wangmeng and Zhang, Lei},
  booktitle={Proceedings of the IEEE conference on computer vision and pattern recognition},
  pages={3262--3271},
  year={2018}
}

@article{zhang2018ffdnet,
  title={FFDNet: Toward a fast and flexible solution for CNN-based image denoising},
  author={Zhang, Kai and Zuo, Wangmeng and Zhang, Lei},
  journal={IEEE Transactions on Image Processing},
  volume={27},
  number={9},
  pages={4608--4622},
  year={2018},
  publisher={IEEE}
}

@article{github,
  title={Github Reposetery - Angular Spectrum Encoding},
  author={Kleiner, Matan and Michaeli, Lior and Michaeli, Tomer},
  journal={https://github.com/matankleiner/Angular-Spectrum-Encoding},
  year={2025}
}
\bibliographystyle{tmlr}

\clearpage

\appendix

\renewcommand\thefigure{S\arabic{figure}}    
\setcounter{figure}{0}  
\renewcommand{\thesection}{\Alph{section}}
\setcounter{section}{0}
\renewcommand{\thetable}{S\arabic{table}}
\setcounter{table}{0}
\renewcommand{\theequation}{S\arabic{equation}}
\setcounter{equation}{0}
\renewcommand{\thesection}{Supplementary Note \arabic{section}}
\renewcommand{\thesubsection}{\arabic{section}.\arabic{subsection}}

\section{Classification}
\label{sm:classification}

We evaluated the proposed angular spectrum encoding method on image classification to illustrate its generality and applicability beyond image-to-image translation. 

\begin{figure}[h]
  \centering
   \includegraphics[width=\linewidth]{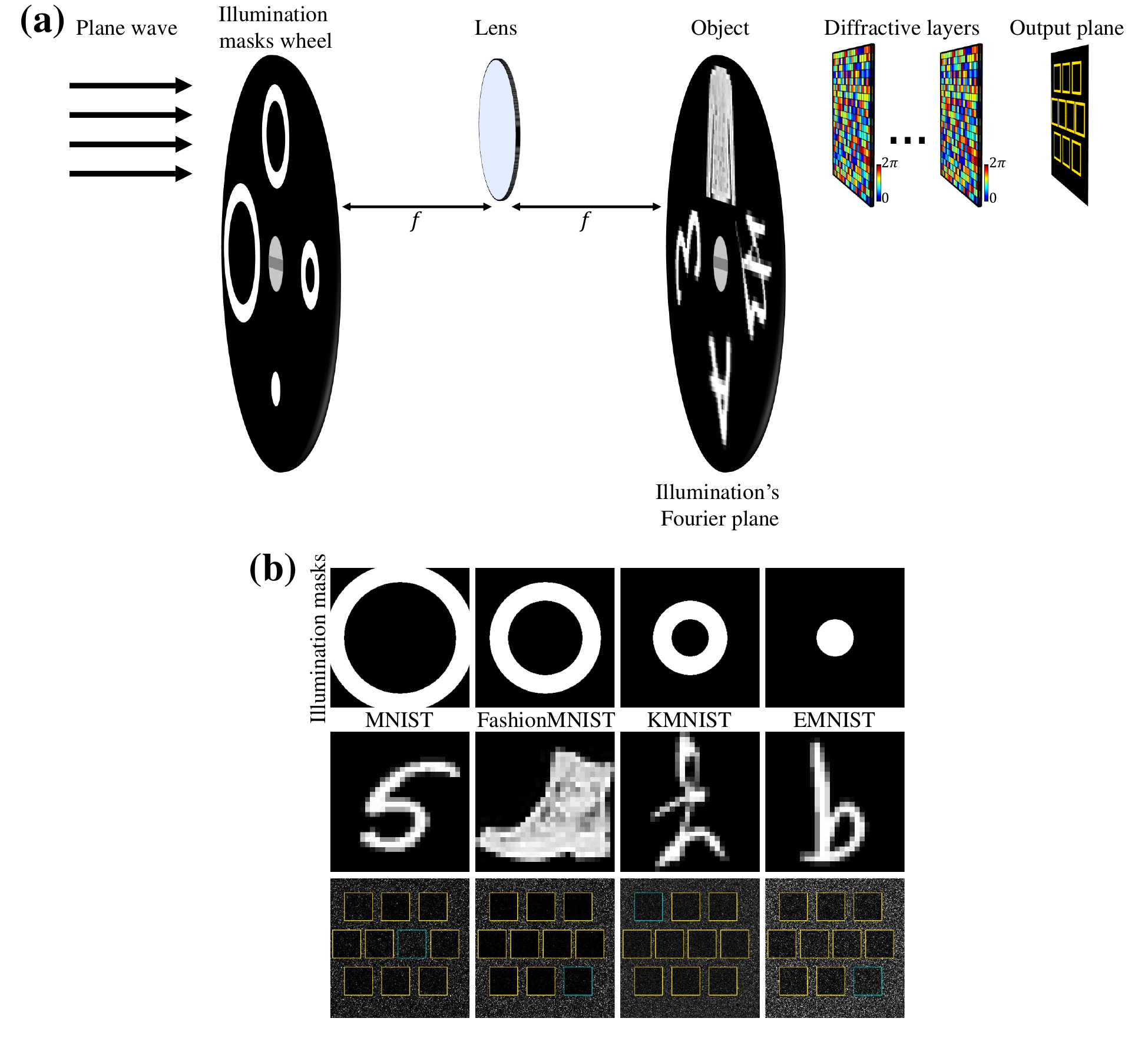}
   \caption{\textbf{Angular spectrum encoding for multi-dataset classification.} (a) Schematic illustration of angular spectrum encoding for object classification. Each illumination mask is used to encode a different dataset. The output plane is comprised of ten predetermined regions, each corresponding to one object class. (b) The output planes of a network trained with angular spectrum encoding for multi-dataset classification. Ring-shaped illumination masks were used for training the network on the MNIST, FashionMNIST, KMNIST and EMNIST datasets.}
   \label{fig:classification}
\end{figure}

Specifically, we consider the setting where a single diffractive network is trained to classify four different datasets, with each illumination mask implicitly encoding a specific datasets. 
The datasets include MNIST (handwritten digits)~\citep{lecun1998mnist}, FashionMNIST (fashion items)~\citep{xiao2017fashion}, KMNIST (handwritten Japanese letters)~\citep{clanuwat2018KMNIST} and EMNIST (handwritten English letters)~\citep{cohen2017emnist}. For all datasets, we used ten classes. The output plane contains ten predefined regions, each corresponding to one class. The output intensity is summed within each region, and the region with the highest total intensity is chosen as the predicted label. Each predefined region is of size $600 \times 600$ $\mu$m. A schematic illustration of angular spectrum encoding for classification is given in Fig.~\ref{fig:classification}a. 

This multi-datasets classification task is fundamentally different from image-to-image translation tasks described in the main text. In image-to-image translation, a single diffractive network produces different output images from the same input image, conditioned on the illumination mask. In contrast, multi-datasets classification involves training each illumination mask with a different dataset, each exhibiting different statistical properties. 

A network trained with angular spectrum encoding using ring-shaped illumination masks achieves classification accuracy of 92.5\%, 81.6\%, 68.2\% and 87.1\% on the MNIST, FashionMNIST, KMNIST and EMNIST datasets, respectively, yielding an average classification accuracy of 82.35\%. The output planes of this network for these different datasets is illustrated in Fig.\ref{fig:classification}b.
The network was trained with four diffractive layers under monochromatic illumination of 550 nm, using the same hyperparameters described in Sec.~\ref{methods}.  
We used 10000 training examples from each dataset and trained the for 1000 epochs, randomly selecting one dataset per epoch. Evaluation was performed on the full test sets, which contain 10000 images for MNIST, FashionMNIST and KMNIST and 48000 images for the EMNIST dataset. 

We consider two baselines for comparison. Both were trained with the same architecture, hyperparameters and training procedure detailed above. 
The first baseline is a network trained without any encoding. This network is implicitly tasked with the challenging problem of identifying the task to solve based on the input. It's classification accuracy is 73.5\% on all datasets. 
The second baseline is a network trained with wavelength encoding, where a distinct wavelength was used for each dataset (400, 500, 600 and 700 nm for MNIST, FashionMNIST, KMNIST and EMNIST, respectively). This network achieves an average classification accuracy of 79.95\%. Both baselines perform worse than the network that was trained with the proposed encoding method, as mentioned above. 

For reference, we also trained four specialized networks under monochromatic illumination of 550 nm, each one trained and evaluated on one dataset.  
The performance drop for the angular spectrum encoding network relative to these specialized networks ranges from 0.7\% to 9.6\%. This drop in performance is the smallest among all multi-task approaches described above. Classification accuracies for all networks are summarized in Tab.~\ref{tab:classification}.

\clearpage

\begin{table}[h]
\caption{\textbf{Classification accuracy of networks trained on four different datasets.} The first row reports the classification accuracy of specialized networks, each trained on and evaluated on a specific dataset. This row provides an upper bound on the performance achievable by the multitask networks in the rest of the rows. The following rows report the classification accuracy of a network trained without any encoding method, with wavelength encoding and with the proposed angular spectrum encoding. All numbers indicate classification accuracy in percentage. Cells colored in \colorbox[HTML]{9AFF99}{green} represent the highest classification accuracy between these three networks.}
\begin{tabular}{@{}cccccc@{}}
\toprule
                      & MNIST                        & FashionMNIST                 & KMNIST                       & EMNIST                       & Avg.~acc.~        \\ \midrule
Specialized               & 93.2                         & 83.5                         & 77.8                         & 88.2                         & -                             \\ \hline \hline
w/o encoding              & 87.5                         & 75.8                         & 56.8                         & 73.9                         & 73.5                          \\
Wavelengths enc.      & 91.5                         & 73.6                         & \cellcolor[HTML]{9AFF99}72.5 & 82.2                         & 79.95                         \\
Angular spectrum enc. & \cellcolor[HTML]{9AFF99}92.5 & \cellcolor[HTML]{9AFF99}81.6 & 68.2                         & \cellcolor[HTML]{9AFF99}87.1 & \cellcolor[HTML]{9AFF99}82.35 \\ \bottomrule
\label{tab:classification}
\end{tabular}
\end{table}

\clearpage
\section{Energy preservation}
\label{sm:energy}

Training a diffractive network using the $\ell_2$ regression loss between the network's outputs, $\hat{I}$, and the ground-truth images, $I_{\text{GT}}$, results in output images that retain $\sim$8\%-1\% of the input energy. 

To improve energy preservation, we modify the training objective to include a combination of a regression loss between the network's \textit{normalized} outputs, $\hat{I}_{\text{norm}}$, and the ground-truth images, $I_{\text{GT}}$, and a loss term that penalizes for low output intensities. 
The normalized output $\hat{I}_{\text{norm}}$ is obtained by dividing $\hat{I}$ by its sum, taken over the entire field of view, and multiplying the result by the sum of intensities of the corresponding ground-truth image. 
For the energy preservation loss, we use an exponential loss function that decreases as the output intensity, $\lVert \hat{I}^{(i,j)} \rVert_1$, increases. 
A hyperparameter $\gamma$ controls the balance between the fidelity and energy preservation loss terms. Thus, the overall training loss is
\begin{equation}\label{eq:loss_energy}
    \mathcal{L} = \frac{1}{N} \sum_{i=1}^N \left(\sum_{j=1}^M \|\hat{I}_{\text{norm}}^{(i,j)} - I_{\text{GT}}^{(c)}\|^2 + \exp\Bigg\{-\gamma\sum_{j=1}^M\| \hat{I}^{(i,j)} \|_1 \Bigg\}\right),
\end{equation}
where $N$ is the number of training examples, $M$ is the number of illumination masks, and $c$ depends on the input image label and the illumination angular components.

\begin{figure}
  \centering
   \includegraphics[width=\linewidth]{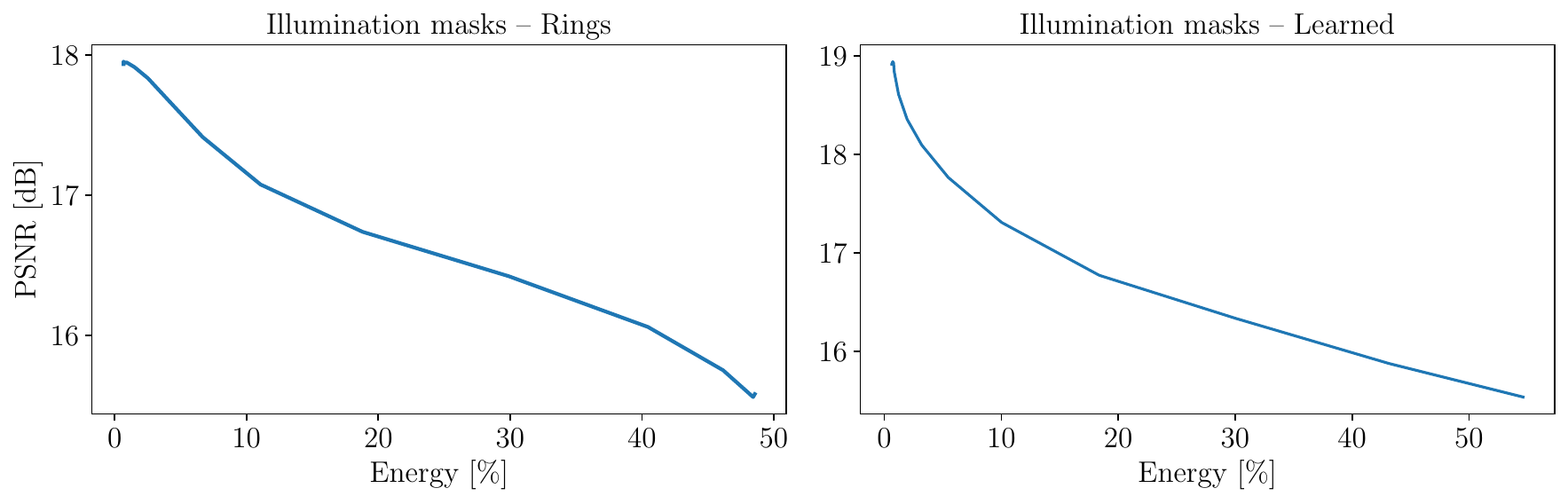}
   \caption{\textbf{Output fidelity vs.~energy preservation.} Results for ring-shaped illumination masks are shown on the left, and results for learned illumination masks are shown on the right.}
   \label{fig:energy}
\end{figure}

Image fidelity is measured by PSNR in dB and energy preservation is measured as the percentage of input energy retained in the output images. 
Figure \ref{fig:energy} illustrate results for networks trained with the same architecture described in the main text but using Eq.~\eqref{eq:loss_energy} as the training objective and with different $\gamma$ values. The left panel shows results with ring-shaped masks and the right panel with learned masks.
The $\gamma$ values for the ring-shaped illumination masks are between $10^{-2}-10^{-7}$ and those for the learned illumination masks are between $10^{-3}-10^{-8}$. All other hyperparameters matched those in the main text, except that training was limited to $100$ epochs, instead of $1000$, to reduce runtime on the expense of a minor performance deterioration. 

\begin{figure}
  \centering
   \includegraphics[width=0.6\linewidth]{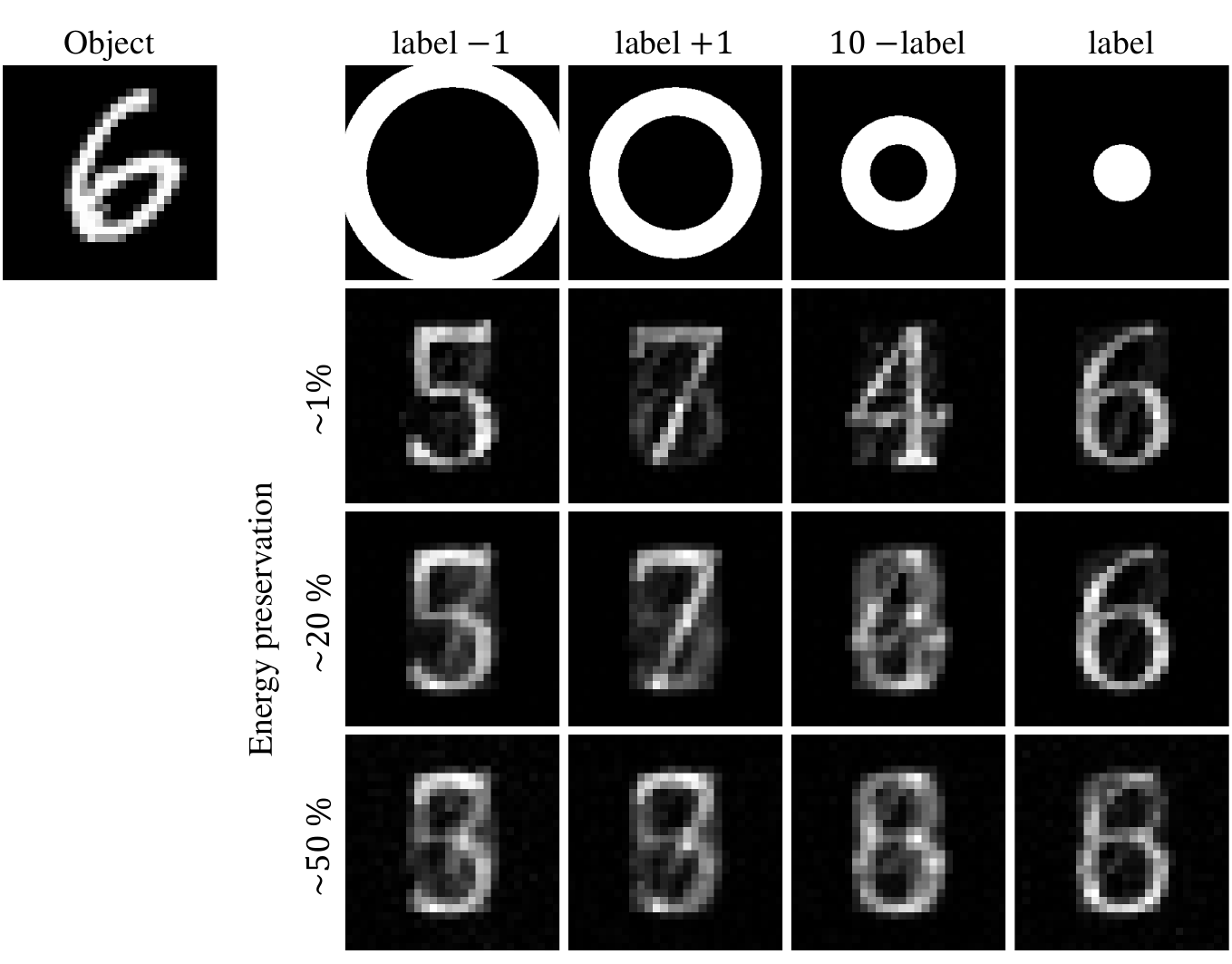}
   \caption{\textbf{Example of ring-shaped illumination masks with different energy preservation.} Energy preservation in percentage is given next to each row of results.}
   \label{fig:energy_res}
\end{figure}

Qualitative results for networks trained with ring-shaped illumination masks using different $\gamma$ values are shown in Fig.~\ref{fig:energy_res}. It can be seen that when the network preserves $\sim 1\%$ of the input energy, it produces sharp and clear results. Higher energy preservation leads to blurrier outputs and occasional failure to produce the correct target image. 

\begin{figure}
  \centering
   \includegraphics[width=\linewidth]{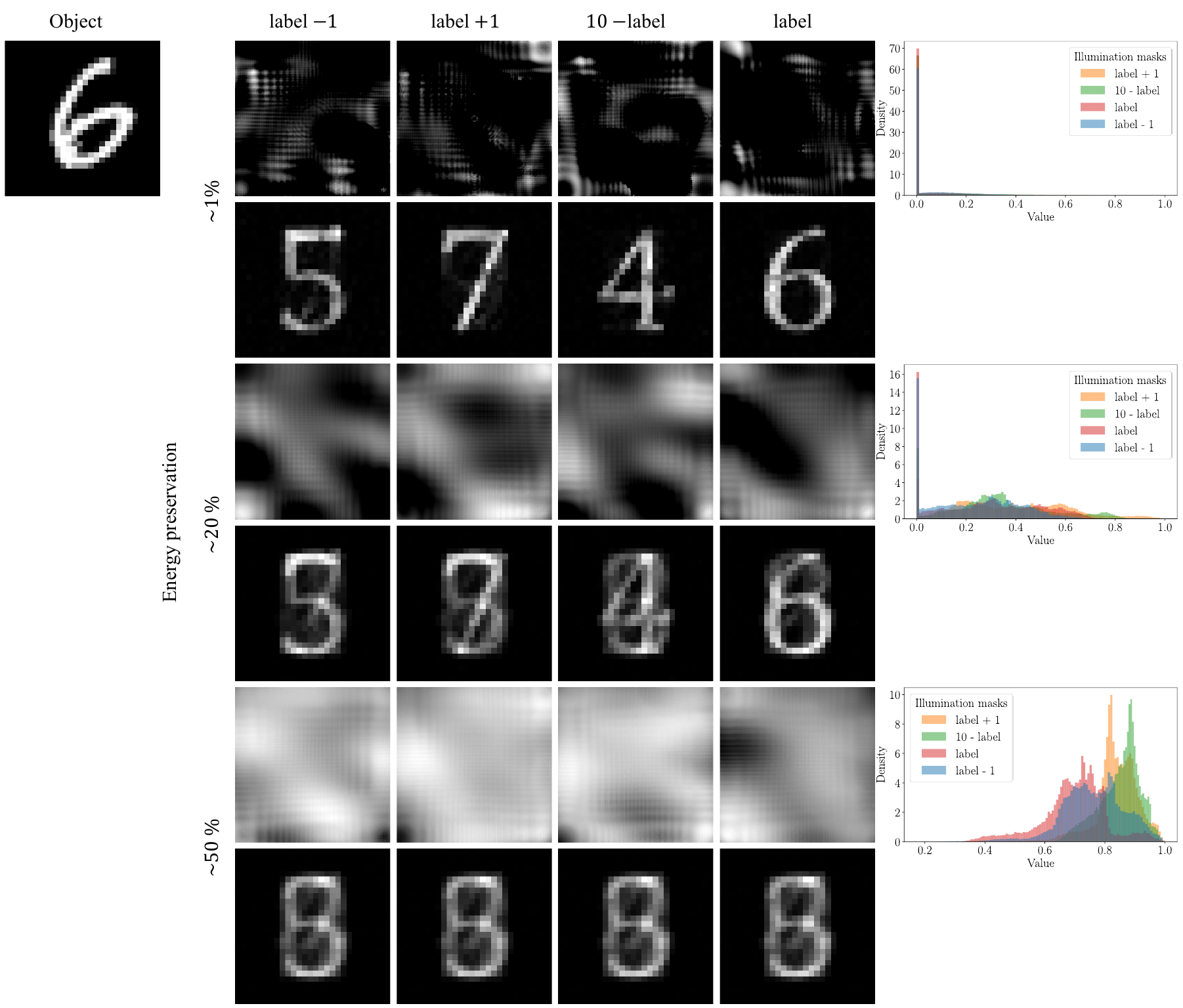}
   \caption{\textbf{Example of learned illumination masks with different energy preservation.} Illumination masks are shown above their corresponding results, alongside a histogram of their pixel values. Energy preservation in percentage is indicated next to each row of results.} 
   \label{fig:energy_res_learned}
\end{figure}

When optimizing illumination masks with Eq.~\eqref{eq:loss_energy} and clamping pixel values to [0,1] (as in the main text), networks tend to either produce high fidelity results or preserve energy, with no clear working point that balances both metrics, across many $\gamma$ values.  
Therefore, we enforced only pixels non-negativity during training and normalized masks to the [0,1] during inference. This approach produced the curve in Fig.~\ref{fig:energy} (right panel) and the qualitative results in Fig.~\ref{fig:energy_res_learned}. 
This training also leads to non-binary illumination masks with many intermediate pixel values, as shown in the histograms next to masks in Fig.~\ref{fig:energy_res_learned}. 
It can be seen that when a trained network preserves $\sim 1 \%$ of the input energy, it produces sharp and clear results, and most of the pixels in the illumination masks are zero or near zero. As energy preservation increases, output fidelity deteriorates and mask pixel values become larger.

\clearpage

\clearpage
\section{Handwritten translation with different illumination masks}
\label{sm:letters}

\begin{figure}[h]
  \centering
   \includegraphics[width=\linewidth]{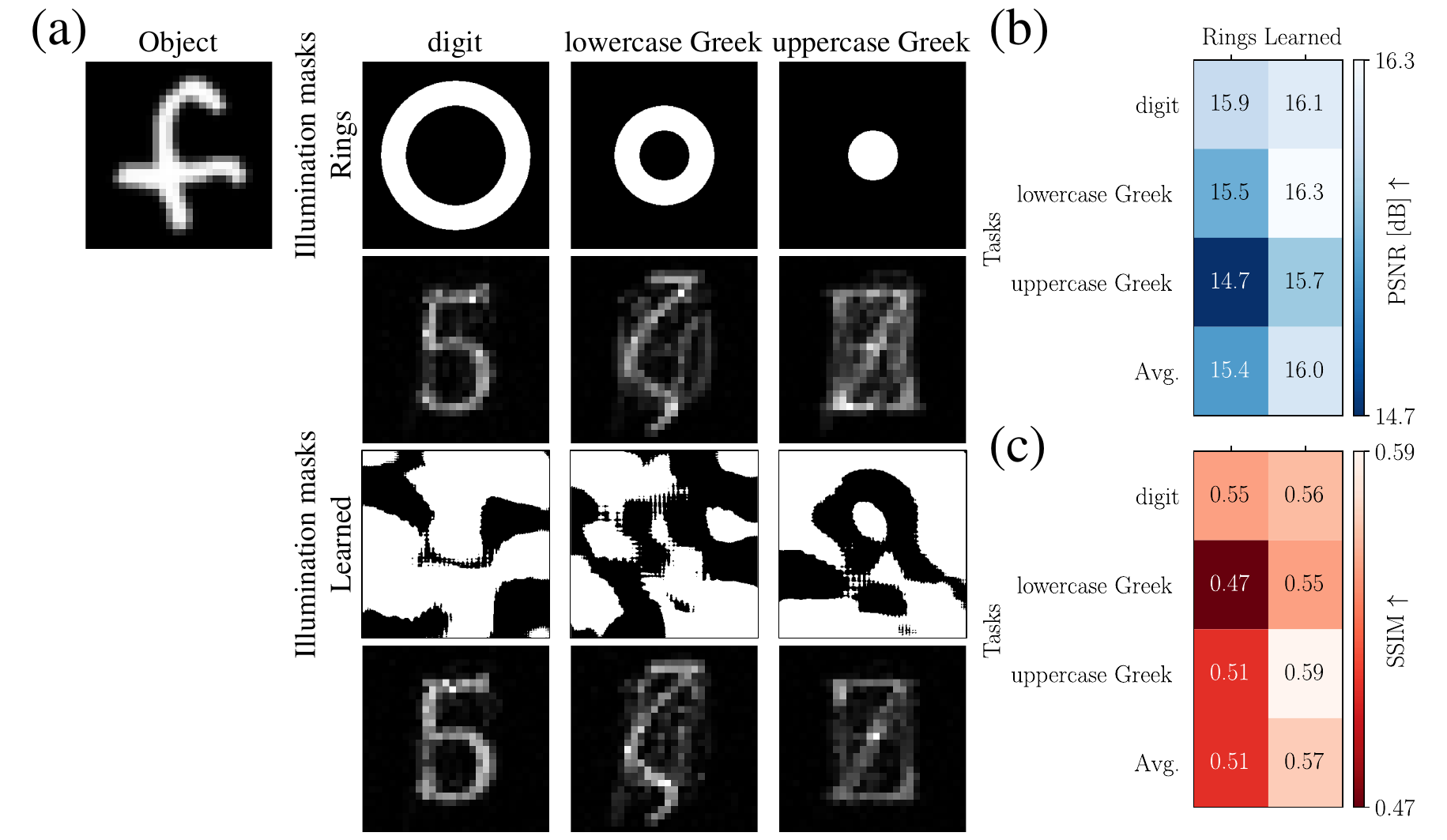}
   \caption{\textbf{Evaluation of angular-spectrum-encoded handwritten letter translation with different illumination masks.} (a) An input image of a handwritten `f' is translated into either a typeset digit or a lowercase Greek letter or an uppercase Greek letters, conditioned on the angular components of the illumination. Rows 1 and 3 show the illumination masks used (Rings and Learned), while rows 2 and 4 show examples for the corresponding network outputs. (b) PSNR and (c) SSIM results averaged over the test set.}
   \label{fig:masks_letters}
\end{figure}

Figure \ref{fig:masks_letters} provides further illustration of the difference between predefined ring-shaped illumination masks and learned illumination masks. Here, handwritten English letters are translated to typeset digits, lower case Greek letters, and uppercase Greek letters. 
Figure \ref{fig:masks_letters}a provides a qualitative example.
Figure \ref{fig:masks_letters}b,c report a quantitative comparison between the ring-shaped and the learned masks.  As in Sec.~\ref{effect} in the main text, the learned illumination masks yield better performance, achieving an average PSNR that is 0.6 dB higher than with the ring illumination masks.

\clearpage
\section{Amplitude and phase of different illumination masks in the illumination's Fourier plane}
\label{sm:amp_phase}

\begin{figure}[h]
  \centering
   \includegraphics[width=0.6\linewidth]{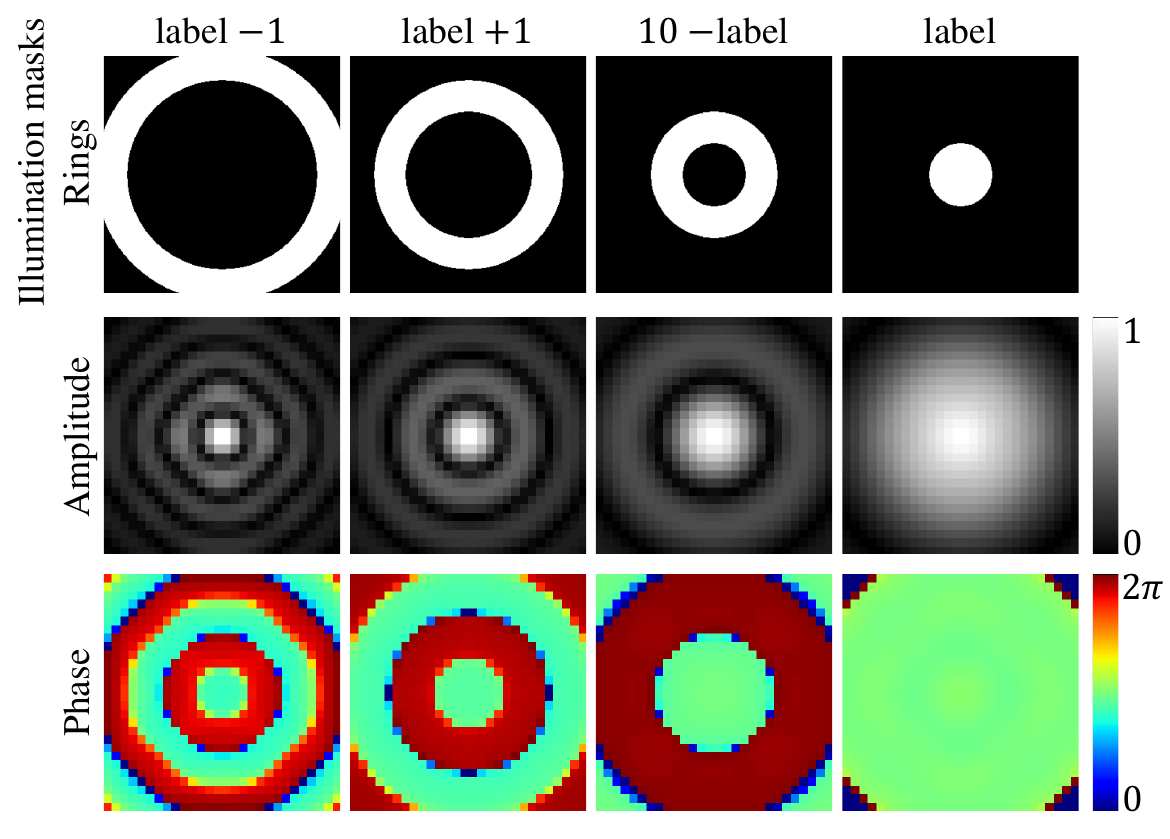}
   \caption{\textbf{The amplitude and phase at the object plane resulting from ring-shaped masks.}}
   \label{fig:amp_phase_r}
\end{figure}

\begin{figure}[h]
  \centering
   \includegraphics[width=0.6\linewidth]{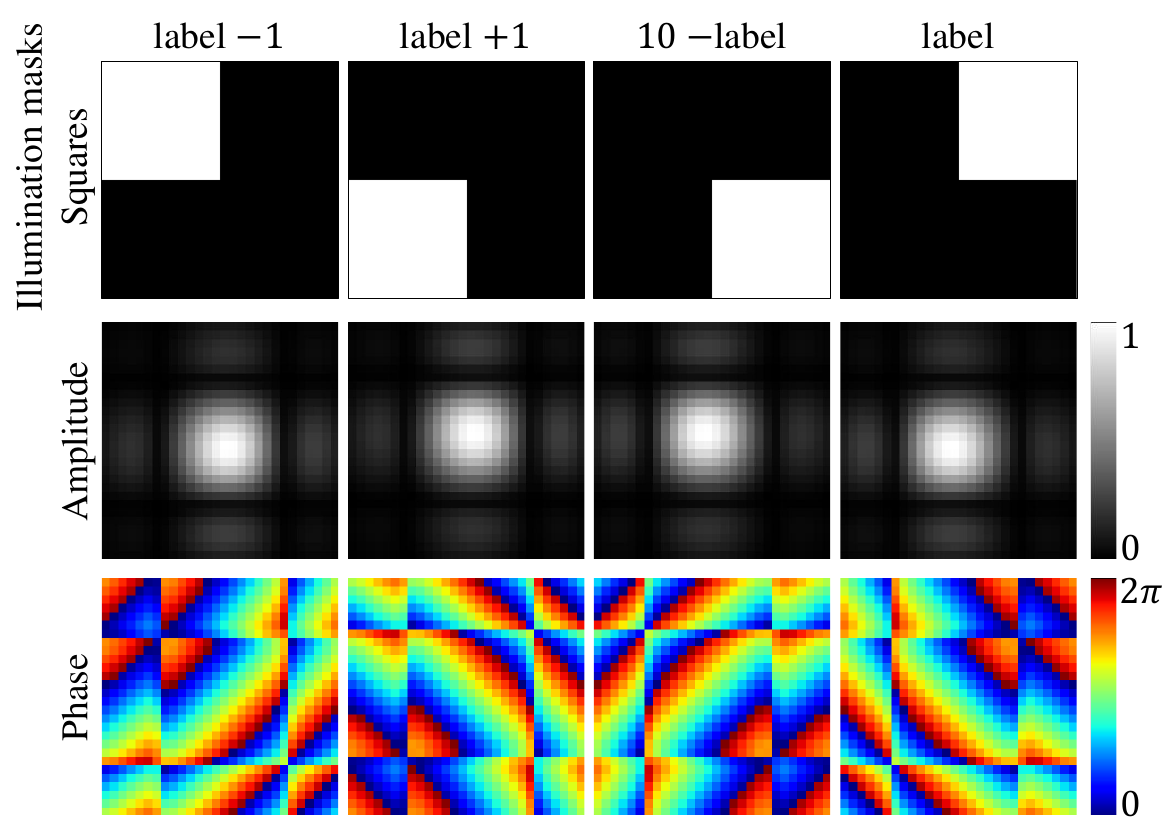}
   \caption{\textbf{The amplitude and phase at the object plane resulting from square-shaped masks.}}   \label{fig:amp_phase_s}
\end{figure}

\begin{figure}[h]
  \centering
   \includegraphics[width=0.6\linewidth]{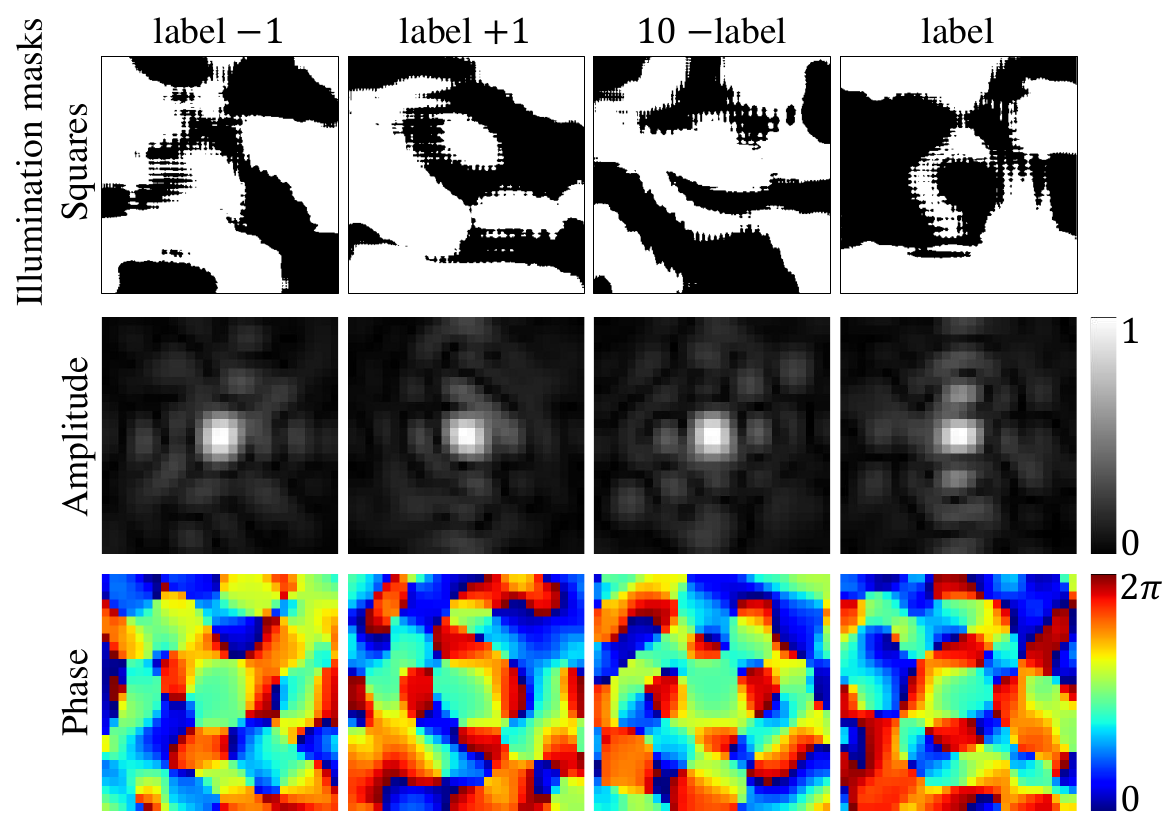}
   \caption{\textbf{The amplitude and phase at the object plane resulting from learned masks.}}   \label{fig:amp_phase_l}
\end{figure}

To gain intuition into the flexibility achievable by our small-NA illumination shaping scheme, figures~\ref{fig:amp_phase_r}),\ref{fig:amp_phase_s}),\ref{fig:amp_phase_l} show the amplitudes and phases at the object plane arising from  the ring-shaped, square-shaped, and learned masks, respectively. 
The amplitude and phases are depicted over the entire input field of view, which is $280 \times 280$ $\mu$m. Amplitude values are normalized to the range $[0,1]$ and phase values are shown between $0-2\pi$.

\clearpage
\section{Combining angular spectrum encoding with wavelength encoding for eight image-to-image translation tasks}
\label{sm:eight}

\begin{figure}[h]
  \centering
   
   \includegraphics[width=\linewidth]{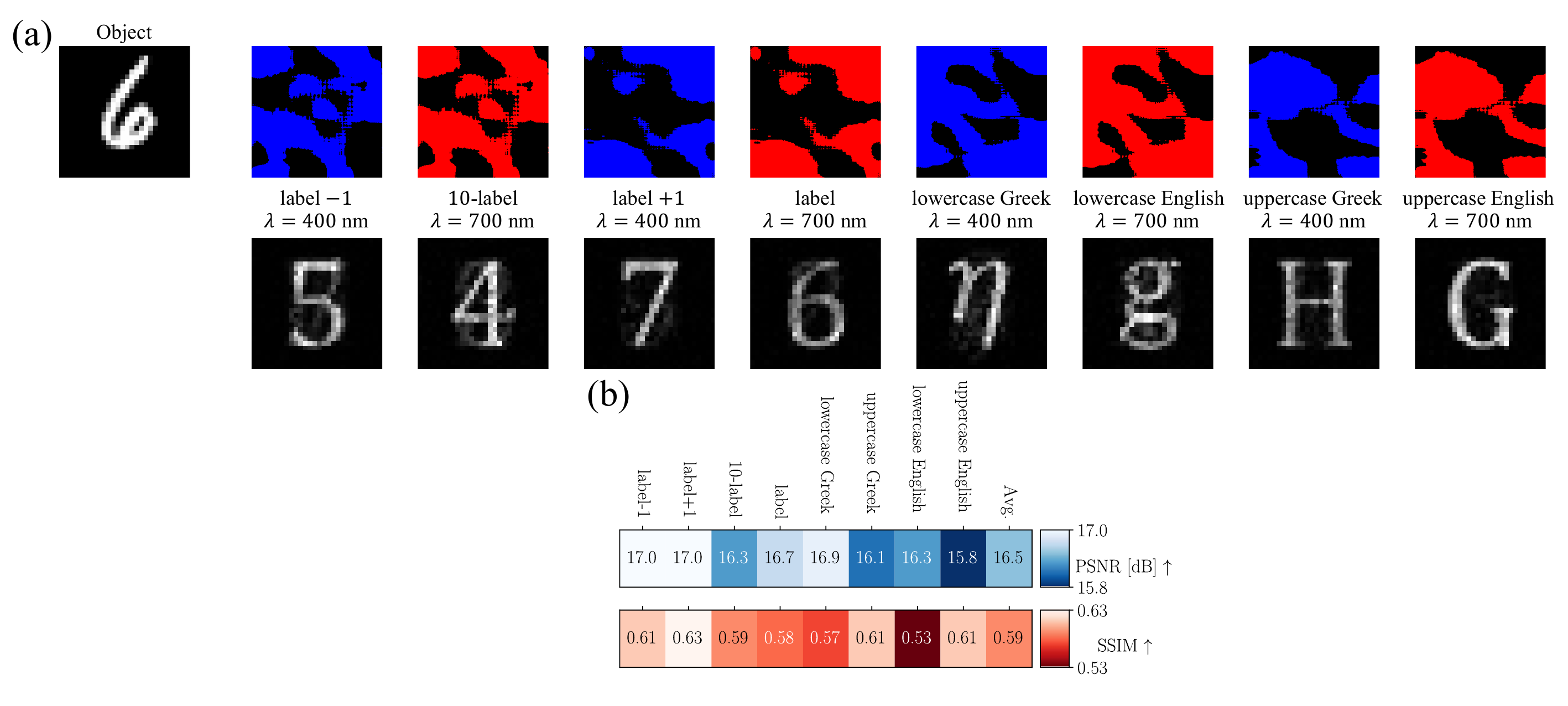} 
   \caption{\textbf{Angular spectrum encoding for handwritten digit translation to eight different target images.} (a) The input, an image of handwritten `6', is translated into typeset digits of different values, Greek and English letters, determined by the angular components of the incident illumination. The network is trained with both angular spectrum encoding and wavelengths encoding (400 and 700 nm), indicated by the colors of the illumination masks. (b) PSNR and SSIM results for the different tasks.} 
   \label{fig:sm_eight}
\end{figure}

Figure~\ref{fig:sm_eight} shows results obtained with a network trained for eight distinct image-to-image translation tasks, by combining angular spectrum encoding with wavelength encoding. The masks used are learned. 
The tasks include producing images of label plus one, label minus one, ten minus the label, the label itself, and lowercase and uppercase letters from both the Greek and English alphabets. The mapping from numbers to letters follows their position in the respective alphabetical index (starting from 0). Combining both encoding methods results in similar performance to the one achieved by using only angular spectrum encoding with learned masks (a 0.1 dB difference in  average PSNR).

\clearpage
\section{Forward propagation through the diffractive network with angular spectrum encoding}
\label{sm:forward}

Simulating propagation of an optical field through a diffractive network with angular spectrum encoding requires simulating free-space propagation and modulation of the optical field by a lens, illumination masks, and the network's diffractive layers. In our implementation the illumination masks are amplitude-only masks and the diffractive layers are phase-only masks.  

Free-space propagation is simulated using the Rayleigh-Sommerfeld diffraction formulation and the angular spectrum method \citep{goodman2005introduction}. The Rayleigh-Sommerfeld transfer function is given by
\begin{equation} \label{eq:RS}  
H_{\text{R-S}}(f_{x},f_{y};z,\lambda) = \begin{cases}
    \exp\left\{j \frac{2\pi}{\lambda} z\sqrt{1-\lambda^2(f_x^2+f_y^2)}\right\} & \text{ if } \sqrt{f_x^2+f_y^2} \leq \frac{1}{\lambda},\\
    0 & \text{otherwise},
\end{cases}
\end{equation}
where $\lambda$ is the wavelength, $z$ is the propagation distance, $f_{x}$ and $f_{y}$ are the spatial frequencies along the $x$ and $y$ directions, respectively. Using this transfer function, we can write the electromagnetic field after propagation by a distance $d$, as
\begin{equation}
    E(x,y;z+d,\lambda) = \mathcal{F}^{-1}\{\mathcal{F}\{E(x',y';z,\lambda)\} \cdot H_{\text{R-S}}(f_{x},f_{y};d,\lambda)\},
\end{equation}
where $\mathcal{F}$ and $\mathcal{F}^{-1}$ are the two-dimensional Fourier transform and inverse Fourier transform operations, respectively.

Modulating the optical field by a thin, amplitude or phase, element, $t$, is simulated as element-wise multiplication (denoted by $\odot$) between the thin element and the optical field just before it $E^-$, given as 
\begin{equation}
    E^+(x,y)=E^-(x,y)\odot t(x,y),
\end{equation}
where $E^+$ is the optical field just after the thin element. 

The illumination mask, $t_{\text{illumination}}(x,y)$, is either predefined or learned. 

The lens is a phase mask with quadratic phase 
\begin{equation}
    t_{\text{lens}}(x,y;\lambda)=\exp\left\{-j\frac{2\pi n}{\lambda}\frac{1}{2f}(x^2+y^2)\right\}.
\end{equation}
where $\lambda$ is the used wavelength, $f$ is the focal length and $n$ is the refractive index. 

The networks' phase masks are simulated as
\begin{equation} \label{eq:phase_modulation}
t_{\text{layer}}(x,y;\lambda) = \exp\{-j\phi(x,y;\lambda)\},   
\end{equation}
where $\phi(x,y;\lambda)$ is the learned phase modulation, which can be written as
\begin{equation}
t_{\text{layer}}(x,y;\lambda) = \exp\left\{-j\frac{2\pi n}{\lambda}h(x,y)\right\},   
\end{equation}
where $h(x,y)$ is a height map. 
The phase modulation for a specific wavelength $\lambda_{0}$ can be written as
\begin{equation} 
t_{\text{layer}}(x,y;\lambda_0)=\exp\left\{-j\frac{2\pi n}{\lambda_{0}}h(x,y)\right\},
\end{equation}
For any different wavelength, $\lambda_{i}\neq\lambda_{0}$, the phase modulation can be written as
\begin{equation} \label{eq:diff_lambda}
t_{\text{layer}}(x,y;\lambda_i)=\exp\left\{-j\frac{\lambda_{0}}{\lambda_{i}}\phi(x,y)\right\}. 
\end{equation}

At the end of the diffractive network, a detector in the output plane measures intensity 
\begin{equation}
I_{\text{spatially coherent}}(x,y;z) = \sum_{\lambda}\left|E(x,y;z,\lambda)\right|^2.
\end{equation}

Spatially incoherent illumination is simulated by multiplying the plane wave with a random phase pattern
\begin{equation}
    t_{\text{random}}(x,y;\lambda_i)=\exp\left\{j\phi(x,y)\right\}, \quad \phi(x,y) \sim \mathcal{U}[0,2\pi].
\end{equation}
This process is repeated $L$ times and intensities are averaged to get the approximate result of the incoherent propagation, 
\begin{equation}
I_{\text{spatially incoherent}}(x,y;z) \approx \frac{1}{L}\sum_{l=1}^L\left|E(x,y;z)\right|^2,
\end{equation}
where the approximation becomes exact as $L\rightarrow \infty$.

\clearpage
\section{Typeset digits and letters}
\label{sm:typeset}

\begin{figure}[h]
  \centering
   \includegraphics[width=\linewidth]{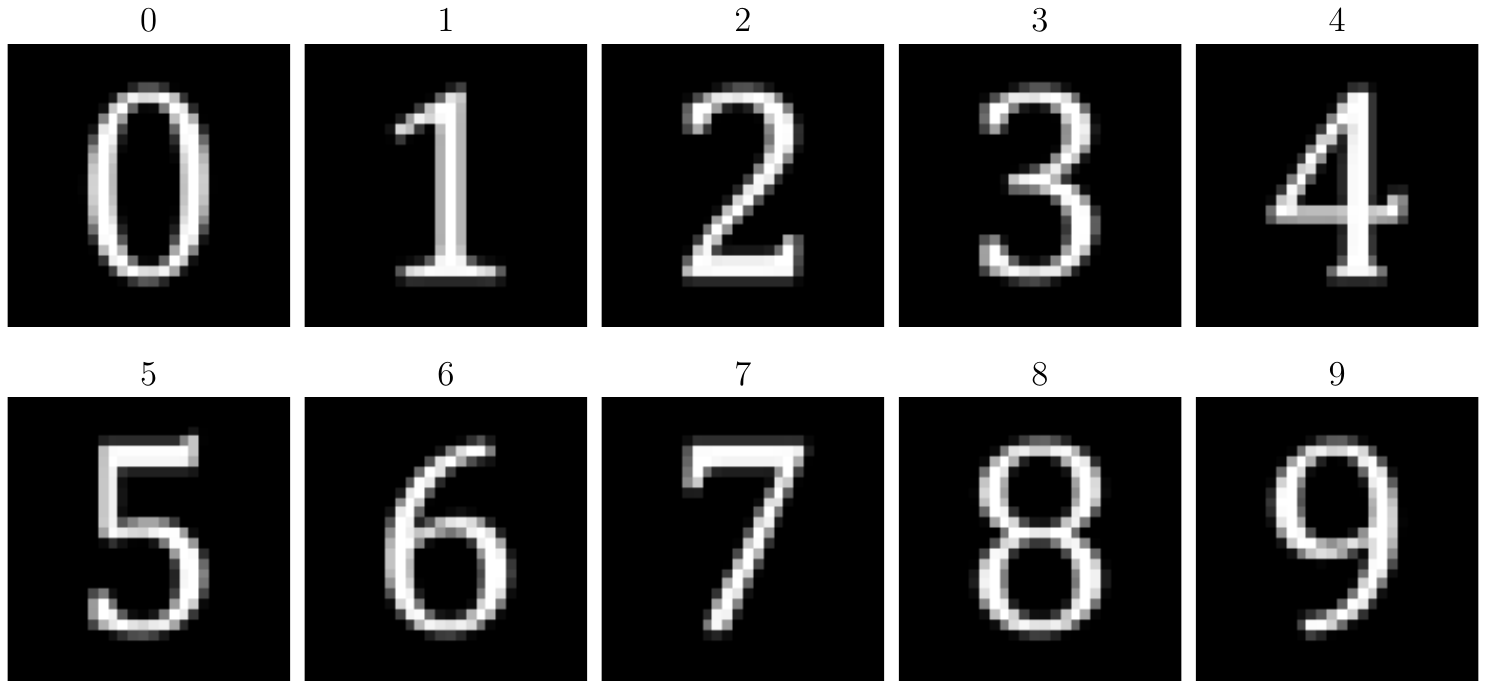}
   \caption{\textbf{Typeset digits.}}
   \label{fig:typeset_n}
\end{figure}

\begin{figure}[h]
  \centering
   \includegraphics[width=\linewidth]{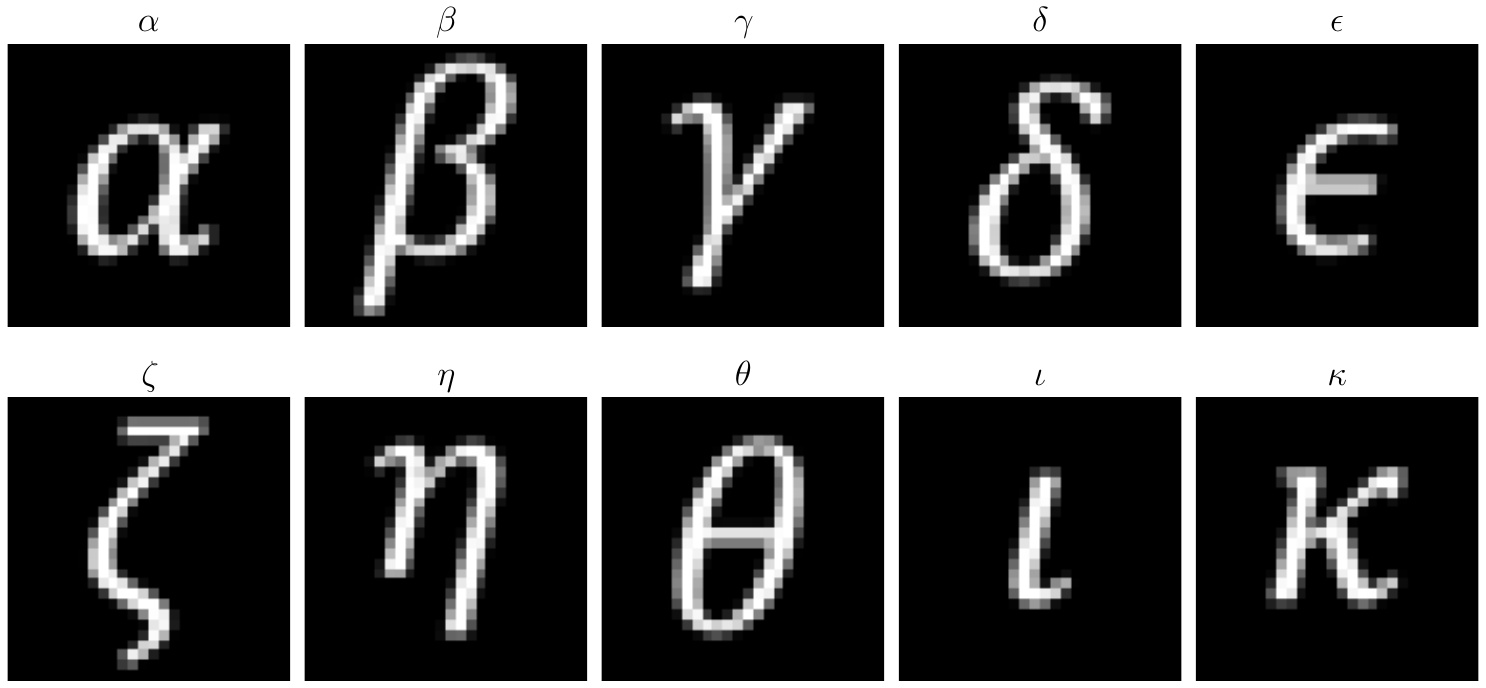}
   \caption{\textbf{Typeset lowercase Greek letters.}}
   \label{fig:typeset_g}
\end{figure}

\begin{figure}[h]
  \centering
   \includegraphics[width=\linewidth]{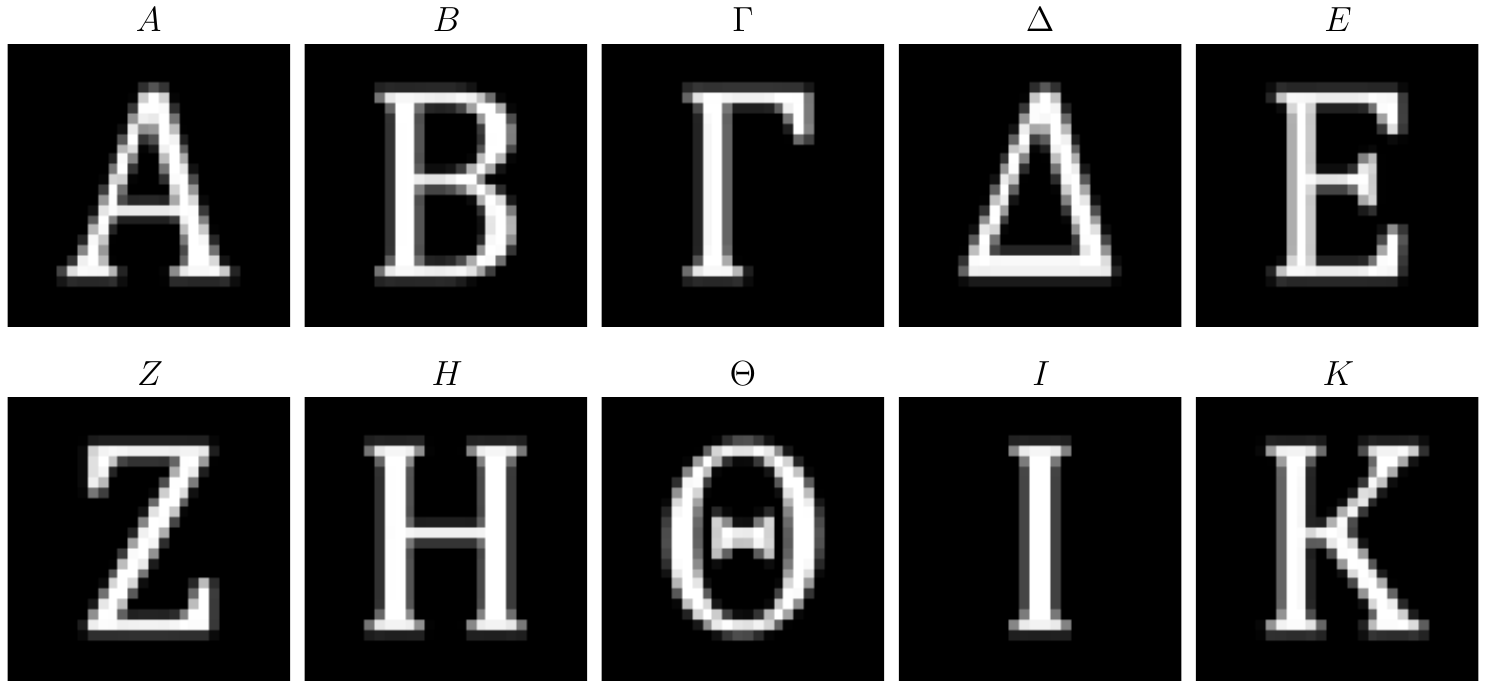}
   \caption{\textbf{Typeset uppercase Greek letters.}}
   \label{fig:typeset_G}
\end{figure}

\begin{figure}[h]
  \centering
   \includegraphics[width=\linewidth]{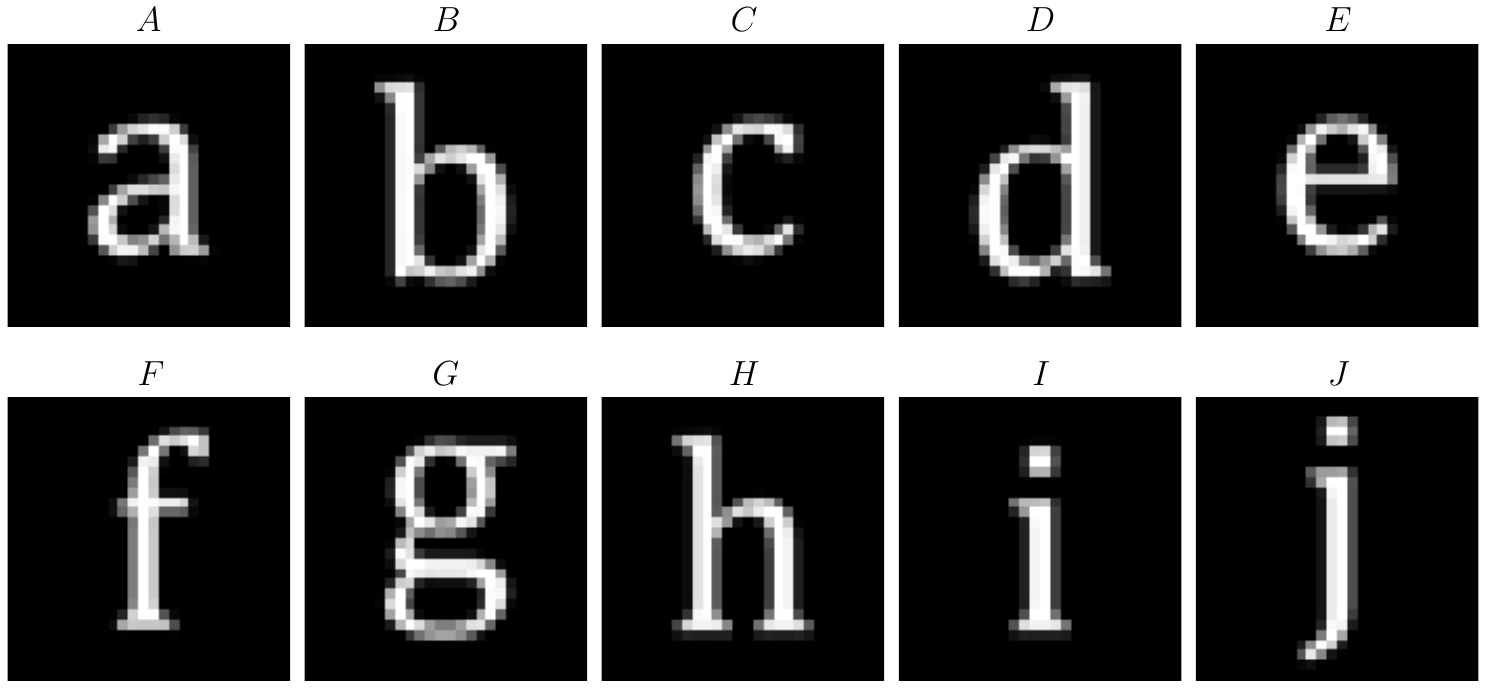}
   \caption{\textbf{Typeset lowercase English letters.}}
   \label{fig:typeset_e}
\end{figure}

\begin{figure}[h]
  \centering
   \includegraphics[width=\linewidth]{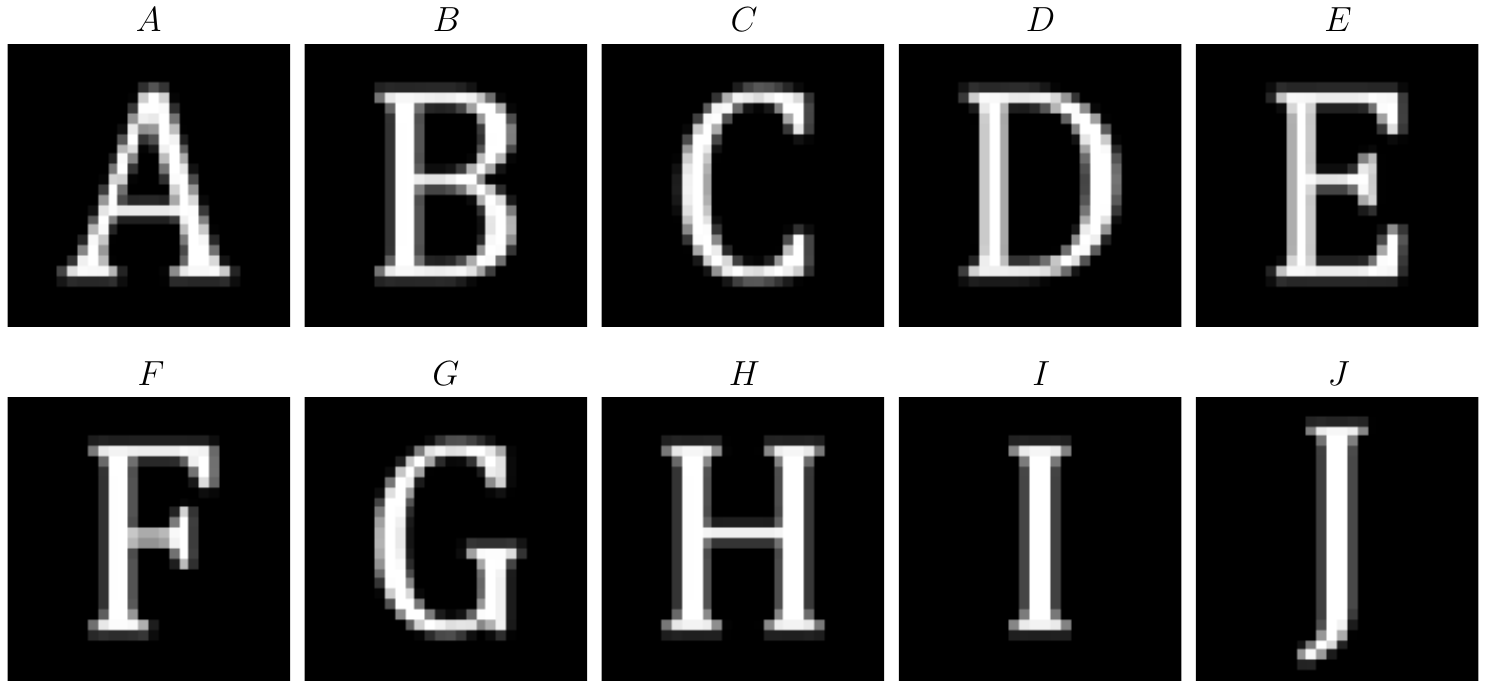}
   \caption{\textbf{Typeset uppercase English letters.}}
   \label{fig:typeset_E}
\end{figure}

The ground-truth images used to train the diffractive networks are provided in Fig.~\ref{fig:typeset_n} (typeset digits), Fig.~\ref{fig:typeset_g} (typeset lowercase Greek letters), Fig.~\ref{fig:typeset_G} (typeset uppercase Greek letters), Fig.~\ref{fig:typeset_e} (typeset lowercase English letters), and Fig.~\ref{fig:typeset_E} (typeset uppercase English letters). All ground-truth images are displayed at a resolution of $28\times28$ pixels, matching the resolution of the input images and the network's output.  

\end{document}